\documentclass[letterpaper,twoside,twocolumn,english,aps,prb,showpacs]{revtex4}
\usepackage[latin1]{inputenc}
\usepackage{graphicx}
\usepackage{amssymb}

\makeatletter

\providecommand{\tabularnewline}{\\}

\usepackage{babel}
\makeatother
\begin{document}

\title{Effects of dephasing on shot-noise in an electronic Mach-Zehnder
interferometer}

\author{Florian Marquardt }

\affiliation{Department of Physics, Yale University, New Haven CT, 06511, USA}

\author{C. Bruder}

\affiliation{Departement Physik und Astronomie, Universität Basel, Klingelbergstr.
82, 4056 Basel, Switzerland}

\date{27.1.2004}

\begin{abstract}
We present a theoretical study of the influence of dephasing on shot
noise in an electronic Mach-Zehnder interferometer. In contrast to
phenomenological approaches, we employ a microscopic model where dephasing
is induced by the fluctuations of a classical potential. This enables
us to treat the influence of the environment's fluctuation spectrum
on the shot noise. We compare against the results obtained from a
simple classical model of incoherent transport, as well as those derived
from the phenomenological dephasing terminal approach, arguing that
the latter runs into a problem when applied to shot noise calculations
for interferometer geometries. From our model, we find two different
limiting regimes: If the fluctuations are slow as compared to the
time-scales set by voltage and temperature, the usual partition noise
expression $\mathcal{T}(1-\mathcal{T})$ is averaged over the fluctuating
phase difference. For the case of {}``fast'' fluctuations, it is
replaced by a more complicated expression involving an average over
transmission amplitudes. The full current noise also contains other
contributions, and we provide a general formula, as well as explicit
expressions and plots for specific examples. 
\end{abstract}


\maketitle

\section{Introduction}

A large part of mesoscopic physics is concerned with exploiting and
analyzing quantum interference effects in micrometer-size electronic
circuits. Therefore, it is important to understand how these interference
effects are diminished by the action of a fluctuating environment
(such as phonons or other electrons), both in order to estimate the
possibilities for applications of these effects as well as for learning
more about the environment itself. This holds for {}``bulk'' effects
such as universal conductance fluctuations and weak localization,
but also for interference in microfabricated electronic interference
setups, such as various versions of {}``double-slit'' or {}``Mach-Zehnder''
interferometers, often employing the Aharonov-Bohm phase due to a
magnetic flux penetrating the interior of the interferometer. In recent
years, many studies\cite{key-11,key-15,key-13,hansen,key-12,key-25,key-28}
have been performed to learn more about the mechanisms of dephasing
and the dependence of the dephasing rate on parameters such as temperature.
One very delicate issue in the analysis of these experiments is the
fact that the {}``visibility'' of the interference pattern can also
be diminished by thermal averaging, when electrons with a spread of
wavelengths (determined by voltage or temperature) contribute to the
current. Recently, an ideal single-channel electronic Mach-Zehnder
interferometer has been realized experimentally for the first time\cite{heiblum}.
The arms of the interferometer have been implemented as edge channels
of a two-dimensional electron gas in the integer quantum hall effect
regime. Besides measuring the current as a function of voltage, temperature
and phase difference between the paths, the authors also measured
the shot noise to be able to distinguish between mere {}``phase averaging''
and genuine dephasing. Although the interpretation of the experimental
results still remains unclear to some extent, the idea of using shot
noise to learn more about dephasing is a very promising one, as it
connects two fundamental topics in mesoscopic physics. 

Most theoretical works on dephasing in mesoscopic interference setups
are concerned with its influence on the average current only (see
Refs.~\onlinecite{aleiner,levinson,seelig,marquardt,marquardtDDD,seelig2}
and references therein). Nevertheless, in some works\cite{qheshotnoise},
the effects of dephasing on shot noise have been studied, employing
the phenomenological {}``dephasing terminal'' approach\cite{buettinelastic,key-9,deJongBeen,vanlangen,qheshotnoise},
where an additional artificial electron reservoir randomizes the phase
of electrons going through the setup. However, this approach does
not include any information about the power spectrum of the fluctuations
in the environment, which, from other studies, is known to play an
important role in discussions of dephasing. Therefore, in the present
work, we have set ourselves the task to analyze the effects of dephasing
on the shot noise in a model that incorporates the fluctuation spectrum. 

Apart from that, the model is deliberately chosen as simple as possible:
The case of a true {}``quantum bath'' will not be treated here.
It is the case relevant for lower temperatures and higher voltages,
when dephasing is produced primarily by spontaneous emission of energy
into the bath. This case immediately leads to a many-body problem
where the Pauli principle plays an important role. Instead, we will
consider a Mach-Zehnder setup (Fig.~\ref{cap:dephsetup}) where the
electrons are subject to the fluctuations of a classical noise field\cite{seelig}
(transmission phases become a Gaussian random process in time). This
describes the case of noisy nonequilibrium radiation impinging on
the system, as well as the effects of the classical part of the noise
spectrum of a quantum-mechanical environment, which should dominate
at higher temperatures. The advantage of a classical noise field is
that we still can use a single-particle picture. We will also compare
and contrast our results with those obtained either from a very simple
classical model of dephasing or the dephasing terminal. Regarding
the dephasing terminal, we will argue that its application to shot
noise in interferometer geometries is most likely plagued with a certain
problem that artificially changes the shot noise result. To the best
of our knowledge, the work presented here is the first microscopic
analysis dealing with the effects of dephasing on shot noise in any
electronic two-way interferometer geometry. This study provides the
basis for dealing with the influence of time-varying potentials on
the shot noise in other two-way interferometer geometries, as well
as for extensions to the case of a true quantum-mechanical environment. 

We will demonstrate that the results depend strongly on whether the
environmental fluctuations are fast or slow with respect to the timescales
set by voltage and temperature. In the {}``slow'' case, the shot
noise merely becomes equal to the phase-average of the usual partition
noise expression\cite{lesovik89}, while the expression for the other
limit is more complicated than that. A brief discussion of a part
of the results has already been presented elsewhere \cite{thePRL}.
Recently, an analysis of dephasing in a mesoscopic resonant-level
detector has been carried out along similar lines, including the effects
on shot noise \cite{aash}.

Our work is organized as follows: After discussing the reduction in
visibility of the current interference pattern (Sec. \ref{sec:Visibility}),
we explain the basic idea behind using shot noise as a tool to distinguish
genuine dephasing from mere phase averaging (Sec. \ref{sec:Shot-noise-as}).
The influence of dephasing on shot noise is then derived both for
a simple classical model (Sec. \ref{sec:Phenomenological-classical-model})
and from the dephasing terminal approach (Sec. \ref{sec:Dephasing-terminal-approach}).
Both of these models are phenomenological, and we explain why we believe
there is a problem with the dephasing terminal approach, as applied
to shot noise calculations in interferometer geometries (Sec. \ref{sec:Possible-shortcoming-of}).
Then we turn to the model of dephasing by a classical fluctuating
potential (Sec. \ref{sec:Dephasing-by-classical}), which permits
to take into account the power spectrum of the environment, in contrast
to the other approaches. We will discuss the general current noise
formula (Eqs. (\ref{eq:generalsplit}),(\ref{eq:wideformula})), as
well as limiting cases (Sec. \ref{sub:Limiting-cases}) and plots
for special examples (Sec. \ref{sub:Current-noise-at}). Finally,
we will compare the results of the various different models and regimes
(Sec. \ref{sec:Comparison-of-different}).

\section{Visibility}

\label{sec:Visibility}The transmission probability (and thus the
current) is determined by squaring the sum of transmission amplitudes
related to the two arms of the interferometer. This results in a sum
of two classical probabilities, plus an interference term depending
on both amplitudes:

\begin{equation}
I\propto|A_{L}|^{2}+|A_{R}|^{2}+A_{L}^{*}A_{R}\left\langle e^{i\delta\varphi}\right\rangle _{\varphi}+c.c.\label{eq:currsimple}\end{equation}
The term $A_{L}^{*}A_{R}$ may contain a fixed Aharonov-Bohm phase
factor, as well as a phase factor $\exp(ik\delta x)$ related to a
possible path-length difference between the two arms. In addition,
there may be an extra fluctuating phase difference $\delta\varphi$,
due to the action of a fluctuating environment, and we have displayed
it explicitly in Eq.~(\ref{eq:currsimple}). In evaluating the current,
one has to perform the average of this fluctuating phase factor $\exp(i\delta\varphi)$,
which will result in a number of magnitude less than one. This decreases
the interference term and therefore the visibility of the {}``interference
pattern'', which is represented by the dependence of $I(\phi)$ on
the controllable Aharonov-Bohm phase difference $\phi$ between the
paths. The visibility is commonly defined as

\begin{equation}
(I_{max}-I_{min})/(I_{max}+I_{min})\,,\end{equation}
where the maximum and minimum of the current is calculated with respect
to $\phi$. This takes on the optimal value of $1$, if the interference
term is not suppressed and the two beam splitters are symmetric ($|A_{L}|=|A_{R}|$).
Any fluctuations in $\delta\varphi$ decrease the visibility. However,
besides temporal fluctuations in $\delta\varphi$, there is another,
more trivial, effect that can diminish the visibility, if there is
a finite path-length difference $\delta x$: This gives rise to an
additional factor $\exp(ik\delta x)$ in the interference term, which
has to be averaged over a range of $k$-values determined by voltage
and temperature (we will call this {}``thermal averaging'', for
brevity). Therefore, in that case the visibility also decreases upon
increasing voltage or temperature.

Therefore, on the level of the average current $I(\phi)$, dephasing
cannot be distinguished easily from thermal averaging. Generically,
even the qualitative dependence of these two different effects on
the parameters ($V,\, T$ etc.) will be similar, since increasing
temperature and/or the bias voltage will usually also increase dephasing.
Nevertheless, in the present model a striking difference appears if
the dependence on the bias voltage is analyzed in more detail: The
average of $\exp(ik\delta x)$ does not simply decrease with increasing
bias voltage, but shows an oscillatory behaviour. Let us illustrate
this briefly in the special case of $T=0$, when the electrons contributing
to the current are injected from the input reservoir in a voltage
window corresponding to a box-distribution of $k$-values ($k\in[\bar{k}-\delta k/2,\bar{k}+\delta k/2]$).
Then we get

\begin{equation}
\left\langle \exp(ik\delta x)\right\rangle _{k}=e^{i\bar{k}\delta x}\frac{\sin(\delta k\delta x/2)}{\delta k\delta x/2}\,,\label{eq:sinx}\end{equation}
which is an oscillatory function that will lead to zero visibility
at all bias voltages for which $\delta k\delta x$ is an integer multiple
of $2\pi$. Finte temperatures will diminish the average further but
will not destroy the oscillatory behaviour. Such an effect, if present,
should be easily confirmed in an experiment. 

However, no such observation has been reported in the Mach-Zehnder
experiment\cite{heiblum}. In general, this absence of oscillations
in the visibility as a function of bias voltage could be taken as
a strong hint for the importance of genuine dephasing, provided our
idealized model applies. In addition, note that there would be a voltage-dependent
phase-shift $\bar{k}\delta x$ of the interference pattern, via Eq.
(\ref{eq:sinx}), which could be used to derive the value of $\delta x$
(provided the Fermi velocity is known), and which could be checked
against the period of the oscillation in the visibility, which is
determined by $\delta x$ as well. Furthermore, it should be noted
that the voltage-dependence of the visibility plotted in Ref.~\onlinecite{heiblum}
was not obtained by simply measuring at different bias voltages. Instead,
a dc-voltage was increased while measuring the ac-current flowing
due to a small ac-modulated voltage on top of the dc bias. Ideally,
the visibility of the ac signal should not decrease with dc-voltage,
if the supression of the interference term were not affected by dephasing
but only by thermal averaging. The actual observation of a decrease
in visibility could therefore be interpreted as another sign ruling
out thermal averaging. Unfortunately, one cannot be sure that the
change of bias voltage does not affect the transmission amplitudes
themselves\cite{heiblumcomm}, and this in turn could mean that the
ac visibility is affected by electron transmission in a wider range
of wavelengths. Thus no firm conclusions can be drawn from the reported
measurements of the voltage dependent visibility.

\section{Shot noise as a measure of dephasing: Basic idea}

\label{sec:Shot-noise-as}Apart from a quantitative analysis of the
temperature- and voltage-dependence of the interference visibility,
there exists another, potentially more powerful way to distinguish
simple thermal averaging from dephasing: shot noise. This was already
pointed out in Ref.~\onlinecite{heiblum}. The basic idea is that
the partition noise $\propto\mathcal{T}(1-\mathcal{T})$ is nonlinear
in the transmission probability $\mathcal{T}$, such that it matters
whether averaging is performed before or after calculating this expression.
Thermal averaging of the independent shot-noise contributions from
different $k$ amounts to an expression of the form

\begin{equation}
\left\langle \mathcal{T}(1-\mathcal{T})\right\rangle _{k}\,.\label{eq:thermalavg}\end{equation}
An analogous expression is expected to hold if some parameter fluctuates
slowly from run to run of the experiment, such that the partition
noise should be averaged over this parameter. In contrast, for the
purposes of interpreting the measurements\cite{heiblum}, dephasing
was assumed to lead to partition noise of the form 

\begin{equation}
\left\langle \mathcal{T}\right\rangle _{\varphi}(1-\left\langle \mathcal{T}\right\rangle _{\varphi})\,,\label{eq:sndeph}\end{equation}
where $\left\langle \mathcal{T}\right\rangle _{\varphi}$ denotes
the transmission probability whose interference term is already suppressed
(partially) due to dephasing.

Even if the visibility turns out to be the same in both cases, the
shot noise expressions (\ref{eq:thermalavg}) and (\ref{eq:sndeph})
are quite different: For example, in the special case of zero visibility
and 50\% transmission of the first beamsplitter ($T_{A}=1/2$), the
shot noise depends on the transmission of the second beam splitter,
$T_{B}$, only in the case of thermal averaging \cite{heiblum}, Eq.
(\ref{eq:thermalavg}). In the case of dephasing, the shot noise in
Eq. (\ref{eq:sndeph}) turns out to be independent of $T_{B}$. In
the next section, we will show that Eq. (\ref{eq:sndeph}) indeed
follows from generalizing the result of a phenomenological classical
model for shot noise of incoherent electrons.

On the other hand, it is clear that the ansatz (\ref{eq:sndeph})
cannot possibly hold in all parameter regimes: In particular, if the
environment-induced fluctuations of the phase are sufficiently slow
(compared with the frequency scale set by temperature or bias voltage,
see below), we would expect that their effect will be just the same
as that of thermal averaging (or that of parameters fluctuating from
run to run of the experiment), leading to a formula similar to Eq.
(\ref{eq:thermalavg}). A qualitative picture is the following: We
may view the current as being composed of a stream of wave packets
entering the interferometer, each of them of a temporal width equal
to the correlation length, i.e. $\textrm{min}(1/k_{B}T,1/eV)$. After
the final beam splitter, the probability weights of the two parts
of the wave packet are determined by the phase difference between
the two interfering paths $L$ and $R$. If the fluctuations of this
phase happen on times much shorter than the temporal extent of the
packet, the probability of detecting the particle in either output
port will be $50/50$ , for \emph{each} packet that enters the interferometer
(in a symmetric setup, with large fluctuations of the phase, leading
to zero visibility). This situation is depicted in Fig.~\ref{cap:fig1}.
On the other hand, if the fluctuations are slow compared to this time
scale, then each packet sent through the setup will feel a fixed (but
random) phase, such that the effects (also in terms of shot noise)
are indistinguishable from thermal averaging. This will be confirmed
by the microscopic model of section \ref{sec:Dephasing-by-classical}.

\section{Phenomenological classical model}

\label{sec:Phenomenological-classical-model}%
\begin{figure}
\begin{center}\includegraphics[%
  width=2.5in,
  keepaspectratio]{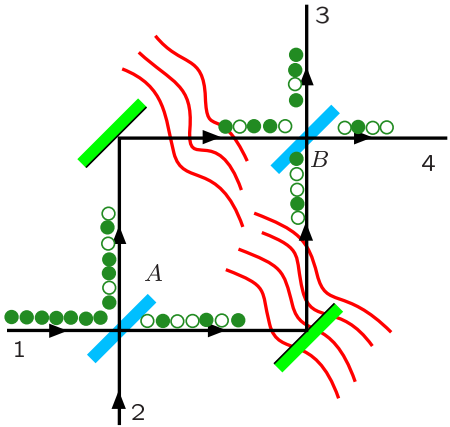}\end{center}

\caption{\label{cap:dephsetup}A simple classical model for the fully incoherent
case: Electrons impinging regularly onto the Mach-Zehnder interferometer
and making classical stochastic choices at both beam splitters.}
\end{figure}

We start from a classical model for shot-noise in a completely incoherent
electronic Mach-Zehnder interferometer setup (see Fig.~\ref{cap:dephsetup}). 

For simplicity, we consider a Mach-Zehnder setup at $T=0$, with a
voltage $V$ applied between the source $1$ and the other terminals.
A heuristic model\cite{key-24} for shot noise calculations consists
in assuming the source emitting electrons in regular intervals of
frequency $eV/h$. It is well-known that this model yields the correct
quantum-mechanical result for the partition noise of a single barrier,
when the variance of the number of transmission events is calculated.
We now go further and implement full loss of phase coherence inside
the interferometer by using classical probability theory to describe
the stochastic choices the electron makes at \emph{each} of two beam
splitters (instead of squaring complex amplitudes describing the coherent
passage through the full device).

Within this model, we can consider the current at the ouput terminal,
$I_{3}$, to be a dichotomous random number ($0$ or $1$), whose
value depends on whether the given electron reaches the output port
$3$ (nothing essential changes for port $4$, other than interchanging
transmission and reflection amplitudes at the second beam splitter).
We obtain

\begin{equation}
\left\langle I_{3}\right\rangle =T_{A}T_{B}+R_{A}R_{B}\equiv\left\langle \mathcal{T}\right\rangle _{\varphi},\,\,\left\langle \delta I_{3}^{2}\right\rangle =\left\langle \mathcal{T}\right\rangle _{\varphi}(1-\left\langle \mathcal{T}\right\rangle _{\varphi})\,,\label{eq:simpleclass}\end{equation}
where $\delta I_{3}=I_{3}-\left\langle I_{3}\right\rangle $ and we
have denoted the fully incoherent transmission probability by $\left\langle \mathcal{T}\right\rangle _{\varphi}$.
The shot noise expression derived from this simple model therefore
agrees (in the fully incoherent limit) with the ansatz considered
in Eq. (\ref{eq:sndeph}). Unfortunately, the generalization to arbitrary
(partial) coherence cannot be made within the present model. Therefore,
we turn to a more sophisticated but still phenomenological approach
which works for any value of the visibility.

\section{Dephasing terminal approach}

\label{sec:Dephasing-terminal-approach}

\newcommand{\p}{\varphi}

\begin{figure}
\begin{center}\includegraphics[%
  width=6cm,
  keepaspectratio]{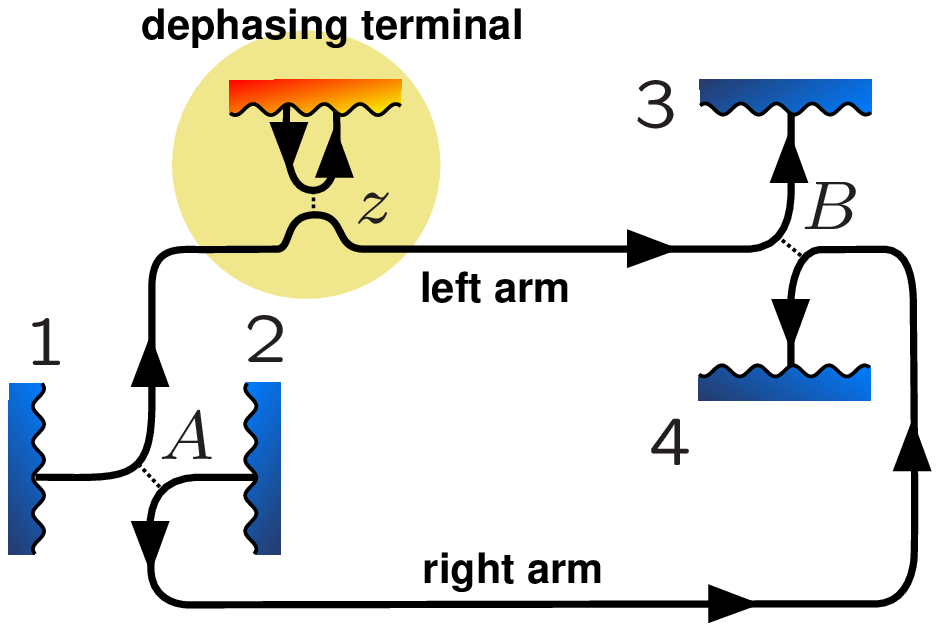}\end{center}

\caption{\label{cap:dephterminalsetup}The Mach-Zehnder interferometer setup
considered in the text: At beam splitters $A$ and $B$ the electrons
are transmitted with amplitudes $t_{A,B}$. The fictitious reservoir
$\varphi$ serves as a {}``dephasing terminal''. The coherence parameter
$z$ denotes the amplitude for an electron to be reflected at the
beam-splitter connecting the left arm of the interferometer to the
reservoir $\varphi$ (thus $z=1$ for fully coherent transport). }
\end{figure}

In this section, we analyze shot noise for a one-channel Mach-Zehnder
setup, employing the dephasing terminal approach\cite{buettinelastic,key-9,deJongBeen,vanlangen,qheshotnoise}.
This will enable us to treat the case of arbitrary visibility, although
it is still not possible to incorporate the spectrum of environmental
fluctuations (see Section \ref{sec:Dephasing-by-classical}). As we
will see, the dephasing terminal model leads to a shot noise expression
which, in general, differs both from $\left\langle \mathcal{T}(1-\mathcal{T})\right\rangle $
and $\left\langle \mathcal{T}\right\rangle (1-\left\langle \mathcal{T}\right\rangle )$. 

Our aim is to calculate the noise of the output current at terminal
$3$ of the interferometer shown in Fig. \ref{cap:dephterminalsetup}.
The basic idea behind the dephasing terminal is to mimick the effects
of dephasing on transport in a mesoscopic conductor by attaching a
fictitious extra reservoir to the setup\cite{buettinelastic,key-9,deJongBeen,vanlangen,qheshotnoise}.
In order to correctly describe pure dephasing, it is essential to
force the current into this dephasing terminal to vanish at each energy
and instant of time. Both the average electron distribution in the
terminal as well as its fluctuations have to be chosen appropriately
to fulfill this condition.

We assume the dephasing terminal $\varphi$ to be attached to the
left interferometer arm (without loss of generality, see below). The
arms are treated as chiral edge channels (see Fig. \ref{cap:dephterminalsetup}).
The amplitude for an electron to move on coherently, without entering
the reservoir, is assumed to be $z\in[0,1]$. When an electron enters
the reservoir with probability $1-z^{2}$, it {}``loses its phase''
and is re-emitted afterwards. In this way, $z$ describes the coherence,
with $z=1$ corresponding to fully coherent transport and $z=0$ to
the completely incoherent case. The amplitude for an electron to go
from reservoir $\beta$ into reservoir $\alpha$ is denoted by the
scattering matrix amplitude $s_{\alpha\beta}$. Assuming backscattering
to be absent at the beamsplitters, the setup of Fig. \ref{cap:dephterminalsetup}
yields the following S-matrix amplitudes: 

\begin{eqnarray}
s_{3\varphi}=ir_{B}e^{i\phi}\sqrt{1-z^{2}} & ,\nonumber \\
s_{31}=t_{A}t_{B}+zr_{A}r_{B}e^{i\phi} & , & s_{\varphi1}=ir_{A}\sqrt{1-z^{2}}\nonumber \\
s_{32}=r_{A}t_{B}+zt_{A}r_{B}e^{i\phi} & , & s_{\varphi2}=it_{A}\sqrt{1-z^{2}}\label{eq:dephtermS}\end{eqnarray}
and $s_{\varphi\varphi}=z$, $s_{33}=s_{\varphi3}=s_{2\varphi}=s_{1\varphi}=0$.
Here $t_{A},r_{A}$ and $t_{B},r_{B}$ are transmission and reflection
amplitudes at the beam splitters $A$ and $B$ (with $|r_{j}|^{2}+|t_{j}|^{2}=1$
and $r_{j}^{*}t_{j}=-r_{j}t_{j}^{*}$). The total phase difference
$\phi$ between the two paths is assumed to include both a possible
Aharanov-Bohm phase, $\phi_{AB}$, as well as the effect of unequal
path lengths, $k\delta x$ (which makes $\phi$ energy-dependent).
Note that, within this model, there is no extra {}``fluctuating phase
difference'' $\delta\varphi$, since dephasing is already included
phenomenologically by the presence of the dephasing terminal.

The current flowing \emph{out} of reservoir $\alpha$ at energy $E$
and time $t$ is given by (note $\hbar\equiv1$)

\begin{equation}
I_{\alpha}(E,t)=\frac{e}{2\pi}(f_{\alpha}-\sum_{\beta}|s_{\alpha\beta}|^{2}f_{\beta})+\delta I_{\alpha}\,,\label{eq:currentalpha}\end{equation}
where $\delta I_{\alpha}$ denotes the original current fluctuations
(at $E,t$) calculated in the absence of any additional fluctuations
of the distribution functions $f_{\alpha}$ (see below). 

Following the calculation of Ref.~\onlinecite{vanlangen}, we demand
the current flowing into the dephasing terminal $\varphi$ to be zero
at each energy and point in time, including its fluctuations. By solving
the equation $I_{\varphi}(E,t)\equiv0$ for $f_{\varphi}$, we obtain:

\begin{equation}
f_{\p}=\left[-\frac{2\pi}{e}\delta I_{\p}+\sum_{\beta\neq\p}|s_{\p\beta}|^{2}f_{\beta}\right]\left[1-|s_{\varphi\varphi}|^{2}\right]^{-1}\,.\label{eq:dephdistr}\end{equation}
The current fluctuations $\delta I_{\varphi}$ on the right hand side
determine the required fluctuations $\delta f_{\varphi}$ of $f_{\varphi}(E,t)=\delta f_{\varphi}(E,t)+\bar{f}_{\varphi}(E)$. 

Inserting $\bar{f}_{\varphi}=R_{A}f_{1}+T_{A}f_{2}$ with $T_{A}=|t_{A}|^{2}$,
$R_{A}=1-T_{A}$ into the averaged Eq. (\ref{eq:currentalpha}), we
obtain the energy-integrated average current at the output port $\alpha=3$:

\begin{equation}
\bar{I}_{3}=\frac{e}{2\pi}\int dE\,\left[f_{3}-f_{1}\left\langle T_{1}\right\rangle -f_{2}\left\langle T_{2}\right\rangle \right]\,,\label{eq:current3}\end{equation}
where the probabilities of transmission from terminals $1$ and $2$
to terminal $3$ are denoted by $\left\langle T_{1}\right\rangle $
and $\left\langle T_{2}\right\rangle $. The notation $\left\langle T_{j}\right\rangle $
is chosen to signal that these transmission probabilities are already
affected by dephasing: They contain an interference term which is
multiplied by the amplitude $z$ of coherent transmission:

\begin{equation}
\left\langle T_{1}\right\rangle =T_{A}T_{B}+R_{A}R_{B}+2z(t_{A}^{*}r_{A})(t_{B}^{*}r_{B})\cos\phi\,,\end{equation}
and $\left\langle T_{2}\right\rangle =1-\left\langle T_{1}\right\rangle $.
For the purposes of calculating the current, the effect of dephasing
may be thought of as an average of the fully coherent expression ($z=1$)
over a fluctuating extra contribution $\delta\varphi$ to the phase
difference $\phi$. This average leads to the suppression of the interference
term: $\left\langle \cos(\phi+\delta\varphi)\right\rangle =z\cos\phi$.
Thus, no simple distinction between genuine dephasing and phase averaging
is possible at this level. The energy-integration in (\ref{eq:current3})
may result in an additional suppression, if there is a difference
in the path lengths of the two interferometer arms (such that $\phi$
is energy-dependent). 

As the phase difference between the two arms is varied (through a
magnetic flux), the current $\bar{I}_{3}$ displays sinusoidal oscillations.
The visibility of this interference pattern, $(I_{max}-I_{min})/(I_{max}+I_{min})$,
is proportional to $z$. If energy averaging is not effective ($\delta k\delta x\ll1$
with $\delta k=\textrm{max}(k_{B}T,eV)/\hbar v_{F}$), the visibility
is equal to $2z\sqrt{T_{A}R_{A}T_{B}R_{B}}/(T_{A}T_{B}+R_{A}R_{B})$.

The \emph{full} current fluctuations $\Delta I_{\alpha}$ at $\alpha\neq\varphi$
contain both the usual fluctuations $\delta I_{\alpha}$, as well
as those induced by the additional fluctuations $\delta f_{\p}$ of
the distribution function in terminal $\varphi$:

\begin{equation}
\Delta I_{\alpha}=\delta I_{\alpha}-\frac{e}{2\pi}|s_{\alpha\p}|^{2}\delta f_{\p}=\delta I_{\alpha}+\frac{|s_{\alpha\p}|^{2}\delta I_{\p}}{1-|s_{\varphi\varphi}|^{2}}\,.\end{equation}
In particular, in our model we obtain

\begin{equation}
\Delta I_{3}=\delta I_{3}+R_{B}\delta I_{\varphi}\end{equation}
for the full current fluctuations at the output port (terminal $3$).
In order to calculate the correlator of $\Delta I_{3}$, we have to
know the correlators of $\delta I_{3}$ and $\delta I_{\varphi}$
(derived for $\delta f_{\p}=0$). According to the scattering theory
of shot noise\cite{buettshotnoise,deJongBeen,key-9,key-24}, we have
in general:

\begin{eqnarray}
P_{\alpha\beta}\equiv2\int dt\,\overline{\delta I_{\alpha}(t+t_{0})\delta I_{\beta}(t_{0})}=\,\nonumber \\
2\frac{e^{2}}{2\pi}\int dE\,\sum_{\gamma,\delta}f_{\gamma}(1-f_{\delta})\nonumber \\
\times(\delta_{\alpha\gamma}\delta_{\alpha\delta}-s_{\alpha\gamma}^{*}s_{\alpha\delta})(\delta_{\beta\gamma}\delta_{\beta\delta}-s_{\beta\delta}^{*}s_{\beta\gamma}).\label{eq:Palphabeta}\end{eqnarray}
The overbar denotes a time-average over $t_{0}$, and the sums run
over all terminals, including the dephasing terminal, where one has
to put $f_{\p}=\bar{f}_{\p}$ for the purposes of this equation. Given
these correlators, we can calculate the noise power at the output
port of the interferometer:

\begin{eqnarray}
2S_{33} & \equiv & 2\int dt\,\overline{\Delta I_{3}(t+t_{0})\Delta I_{3}(t_{0})}=\nonumber \\
 &  & P_{33}+2R_{B}\, P_{3\varphi}+R_{B}^{2}P_{\varphi\varphi}.\label{eq:shotnoisegeneral}\end{eqnarray}

For simplicity, we first focus on the special case of zero temperature
and no path-length difference ($\phi$ energy-independent). A bias
voltage $V$ is applied between terminal $1$ and the other terminals:
$f_{1}(E)=\theta(\epsilon_{F}+eV-E)$, $f_{2}(E)=f_{3}(E)=\theta(\epsilon_{F}-E)$.
From (\ref{eq:Palphabeta}), (\ref{eq:shotnoisegeneral}) and the
scattering matrix amplitudes, we find :

\begin{equation}
\left(\frac{e^{3}V}{2\pi}\right)^{-1}S_{33}=\left\langle T_{1}\right\rangle \left\langle T_{2}\right\rangle -2(1-z^{2})R_{A}R_{B}T_{A}T_{B}\,.\label{eq:shotnoise}\end{equation}
Apparently, even for the fully incoherent case $z=0$ the shot noise
is \emph{not} given by the simple expression $\left\langle T_{1}\right\rangle \left\langle T_{2}\right\rangle =\left\langle T_{1}\right\rangle (1-\left\langle T_{1}\right\rangle )$,
involving the product of averaged transmission probabilities (contrary
to the result of the simple classical model in the previous section).
However, it is interesting to note that this expression would indeed
be found if one were to demand only the \emph{average} current into
the dephasing terminal to vanish at each energy, while we have also
taken into account the restriction for the current fluctuations themselves.
We will comment further on this difference between the two models
in the next section. For the remainder of this section, we will just
discuss the consequences of Eqs. (\ref{eq:shotnoisegeneral}) and
(\ref{eq:shotnoise}).

We note that the result is independent of the location of the dephasing
terminal: Indeed, placing the terminal into the right arm amounts
to the replacements $\phi\mapsto-\phi,\, t_{A}\leftrightarrow r_{A},\, t_{B}\leftrightarrow r_{B}$,
which leave Eq. (\ref{eq:shotnoise}) invariant. More generally, repeating
the analysis with a dephasing terminal in each arm gives exactly the
same results as before, with $z=z_{L}z_{R}$ the product of the amplitudes
for coherent transmission in each arm. Physically this is to be expected,
since the effect of dephasing is only to scramble the \emph{relative}
phase between the two paths. 

\begin{figure}
\includegraphics[%
  width=6cm,
  keepaspectratio]{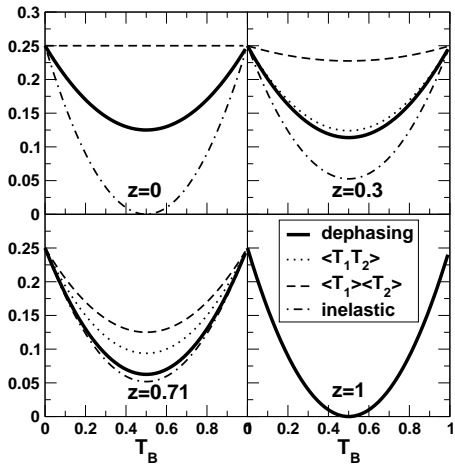}

\caption{\label{cap:plotnoise}Normalized noise power $S_{33}/(e^{3}V/2\pi)$
vs. transmission probability $T_{B}$ of second beamsplitter: Pure
dephasing (Eq. (\ref{eq:shotnoise}), thick line) compared with the
phase-averaged partition noise expression $\left\langle T_{1}T_{2}\right\rangle $
($T_{2}=1-T_{1}$), the product of phase-averaged probabilities, $\left\langle T_{1}\right\rangle \left\langle T_{2}\right\rangle $,
and inelastic scattering (Eq. (\ref{eq:inelastic}), symmetric case
$\lambda=1/2$). Different panels show various values of the coherence
parameter $z$, with a maximum deviation between pure dephasing and
the phase averaged result at $z=1/\sqrt{2}\approx0.7$. Other parameters:
$T_{A}=1/2,\,\phi=0$.}
\end{figure}

In order to compare expression (\ref{eq:shotnoise}) to the result
of a phase-averaged partition noise expression, $\left\langle T_{1}T_{2}\right\rangle $,
we have to evaluate $\left\langle \cos^{2}(\phi+\delta\phi)\right\rangle =(1+\left\langle \cos(2(\phi+\delta\phi)\right\rangle )/2$.
This is not simply related to $z$ (which has been defined via the
average of $\cos(\phi+\delta\phi)$). However, if we assume the phase
fluctuations $\delta\phi$ to be Gaussian distributed, then $\left\langle \cos(2(\phi+\delta\phi)\right\rangle =z^{4}\cos(2\phi)$.
In that case we obtain:

\begin{eqnarray}
\left\langle T_{1}T_{2}\right\rangle -\left\langle T_{1}\right\rangle \left\langle T_{2}\right\rangle =\left\langle T_{1}\right\rangle ^{2}-\left\langle T_{1}^{2}\right\rangle  & =\nonumber \\
4R_{A}R_{B}T_{A}T_{B}(\left\langle \cos\phi\right\rangle ^{2}-\left\langle \cos^{2}\phi\right\rangle ) & =\nonumber \\
-2R_{A}R_{B}T_{A}T_{B}(1-z^{2})(1-z^{2}\cos(2\phi))\label{eq:t1t2avg}\end{eqnarray}
We conclude that for zero visibility ($z=0$) the shot noise expression
(\ref{eq:shotnoise}) is equal to $\left\langle T_{1}T_{2}\right\rangle $,
i.e. it has the form expected from a simple phase average! Therefore,
according to the dephasing terminal approach, in this particular limit
a shot noise measurement could not be used to distinguish phase averaging
and genuine dephasing. 

The coincidence between phase averaging and dephasing holds only at
$z=0$ (and, trivially, at $z=1$). The difference between $\left\langle T_{1}T_{2}\right\rangle $
and the expression given in Eq. (\ref{eq:shotnoise}) is maximized
if $T_{A}=T_{B}=1/2$, $\phi=0,\pi,2\pi,\ldots$ and $z^{2}=1/2$.
At these parameter values, the shot noise expression is $30\%$ below
the value of $\left\langle T_{1}T_{2}\right\rangle $, see Fig.~\ref{cap:plotnoise}.

If phase averaging (against which the pure dephasing case is to be
compared) is actually due to energy integration over a phase factor
$\exp(ik\delta x)$, then the distribution of $\delta\phi$ is not
Gaussian but determined by voltage and temperature. In that case,
we define a parameter $z_{4}$ by $\left\langle \cos(2(\phi+\delta\phi))\right\rangle =z_{4}\cos(2\phi)$.
Here it is understood that $\left\langle \delta\phi\right\rangle =0$
(so $\phi$ corresponds to the average phase), and we have $z_{4}=z^{4}$
for the Gaussian case. In Eq. (\ref{eq:t1t2avg}) the factor $z^{2}$
in front of $\cos(2\phi)$ changes to $(z_{4}-z^{2})/(z^{2}-1)$.
For example, at $T=0$ we have to average over a box distribution
of width $\delta k=eV/(\hbar v_{F})$, which yields $z=\sin(\delta k\delta x)/(\delta k\delta x)$and
$z_{4}=2\sin(\delta k\delta x/2)/(\delta k\delta x)$ (compare the
discussion in Section \ref{sec:Visibility}). Hence the phase-averaged
shot noise $\left\langle T_{1}T_{2}\right\rangle $ can still depend
on the average phase $\phi$ even when the visibility is zero ($z=0,\, z_{4}\neq0$
for $\delta k\delta x=(2n+1)\pi$), in marked contrast to dephasing
or Gaussian phase fluctuations.

If we use the extra terminal $\varphi$ to model inelastic relaxation\cite{buettinelastic}
instead of pure dephasing, its distribution function $f_{\varphi}$
is given by an equilibrium Fermi function of appropriate chemical
potential, and the only condition is that the energy-integrated current
must vanish at each instant of time (voltage probe). This implies
that the chemical potential at this reservoir fluctuates. It turns
out that in the inelastic case it does matter whether relaxation is
ascribed fully to one arm or to both arms. Therefore, we set up a
model with reservoirs $L,\, R$ with associated amplitudes $z_{L},z_{R}$.
As the current only depends on $z_{L}z_{R}\equiv z$, we write $z_{L}=z^{\lambda}$
and $z_{R}=z^{1-\lambda}$, where the parameter $\lambda$ quantifies
the asymmetry ($\lambda=1,0$ for relaxation in the left/right arm
and $\lambda=1/2$ for the symmetric case). In evaluating the shot
noise at terminal $3$ we have to take into account the current correlations
between terminals $3,\, L$ and $R$, along the same lines as before.
The expression in Eq. (\ref{eq:shotnoise}) is replaced by:

\begin{equation}
\left\langle T_{1}\right\rangle \left\langle T_{2}\right\rangle -2R_{A}T_{B}R_{B}(1+(1-2T_{A})z^{2}-R_{A}(z^{2(1-\lambda)}+z^{2\lambda})),\label{eq:inelastic}\end{equation}
for $R_{A}<T_{A}$ (otherwise interchange $R_{A},T_{A}$). In the
fully asymmetric case ($\lambda=0,1$), we recover the result (\ref{eq:shotnoise})
obtained for pure dephasing. However, in general the shot noise may
be reduced: For example, at $\lambda=1/2$ and $T_{A}=1/2$ Eq. (\ref{eq:inelastic})
turns into $\left\langle T_{1}\right\rangle \left\langle T_{2}\right\rangle -R_{B}T_{B}(1-z)$,
which can become zero even in the limit of full relaxation ($z=0$),
at $T_{B}=1/2$ (see Fig. \ref{cap:plotnoise}). 

For reference purposes, we also list the generalization of the pure
dephasing result, Eq. (\ref{eq:shotnoise}), to the case of finite
temperatures and energy-dependent transmission probabilities:

\begin{eqnarray}
\frac{2\pi S_{33}}{e^{2}}=\int dE\,(\delta f\left\langle T_{1}\right\rangle +f)(1-(\delta f\left\langle T_{1}\right\rangle +f))\nonumber \\
+f(1-f)-2(1-z^{2})R_{A}R_{B}T_{A}T_{B}\delta f^{2}\end{eqnarray}
Here $f=f_{2}=f_{3}$ is a thermally smeared Fermi function, and $\delta f(E)=f(E-eV)-f(E)$
is the difference of distributions in reservoirs $1$ and $2$. 

It should be emphasized that the phenomenological dephasing terminal
approach cannot yield the dependence of dephasing strength on voltage
and temperature, since the strength of dephasing, $z$, enters as
an arbitrary parameter. It also does not account for the spectrum
of environmental fluctuations, which is important to provide a smooth
cross-over between dephasing and phase averaging if the ratio of the
typical fluctuation frequencies to other characteristic frequencies
(e.g. voltage and temperature) is varied (see Section \ref{sec:Dephasing-by-classical}). 

In this section, we have calculated the effect of pure dephasing on
shot noise in a mesoscopic Mach-Zehnder interferometer for electrons,
using the scattering theory of shot noise and the dephasing terminal
approach. The resulting shot noise expression is, in general, different
from a simple phase average of the usual partition noise formula,
and therefore may be employed to distinguish real dephasing from mere
phase averaging. However, the result also differs from what one might
obtain by merely inserting transmission probabilities where the effect
of dephasing has already been taken into account. We have pointed
out that dephasing and phase averaging become indistinguishable in
the limit of zero visibility (within this model), but that a strong
difference may be observed for other parameter values.

\section{Possible shortcoming of the dephasing terminal}

\label{sec:Possible-shortcoming-of}

In this section, we reexamine the difference between the shot noise
results obtained from the simple classical model of Section \ref{sec:Phenomenological-classical-model}
and the dephasing terminal of the previous section. We will take the
classical model as our starting point and investigate how the dephasing
terminal approach would be implemented within the context of this
model. As we will see, the extra suppression of shot noise in the
dephasing terminal turns out to be artificial. 

\begin{center}%
\begin{figure}
\begin{center}\includegraphics[%
  width=0.45\columnwidth,
  keepaspectratio]{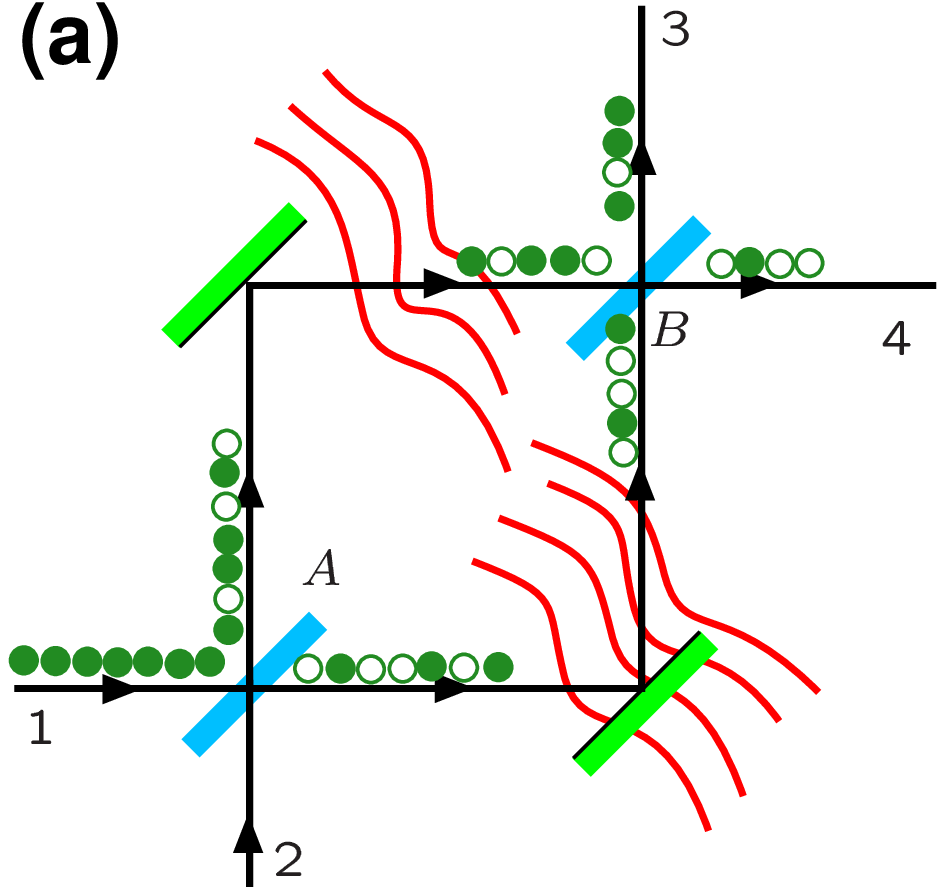}~\includegraphics[%
  width=0.48\columnwidth,
  keepaspectratio]{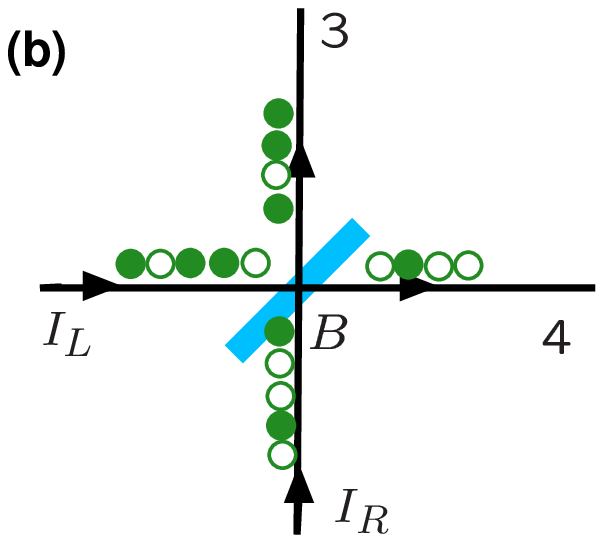}\\
\includegraphics[%
  width=0.48\columnwidth]{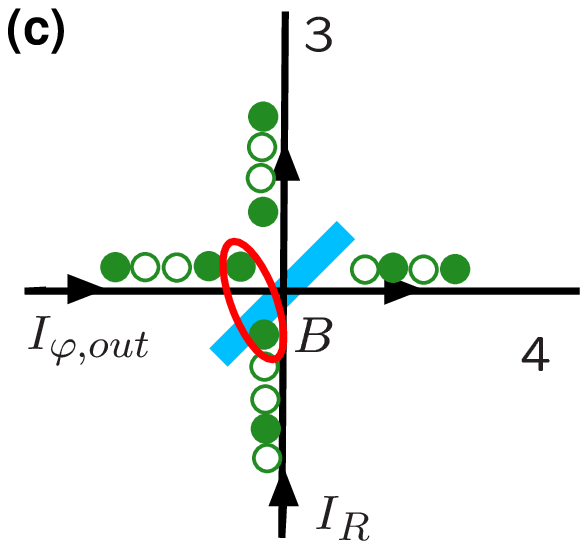}~\includegraphics[%
  width=0.45\columnwidth,
  keepaspectratio]{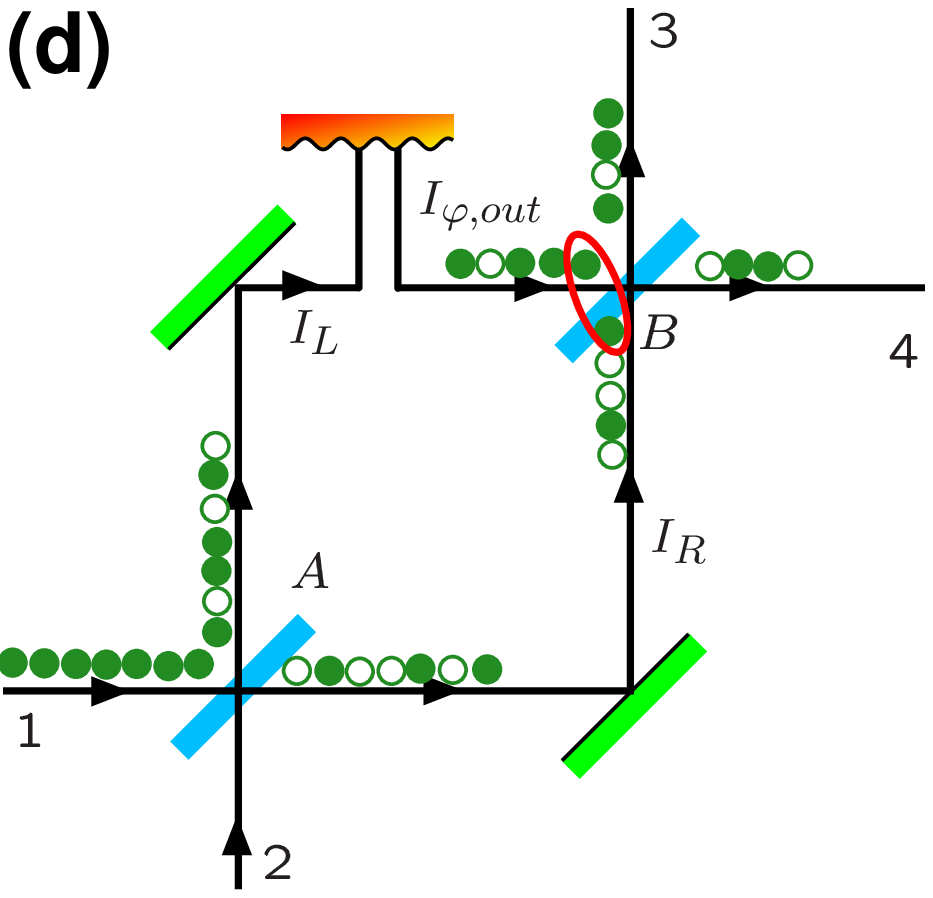}\end{center}

\caption{\label{cap:dephclassical}Interpretation of the dephasing terminal
within the context of the simple classical model (a): The shot-noise
reduction due to anticorrelations (b) and the Pauli principle (c)
are kept at the same time in the dephasing terminal approach (d).}
\end{figure}
\end{center}

For simplicity, we restrict attention to the case of a symmetric first
beam splitter, $T_{A}=1/2$. We now focus on the second beam splitter
$B$ and ask for the shot noise at its output port. The initial classical
model (see Fig. \ref{cap:dephclassical}a) leads to perfectly anticorrelated
streams of electrons in the left and right arm, entering beam splitter
$B$ (see Fig. \ref{cap:dephclassical}b). Thus we can obtain the
correct result by treating the inputs as two incoherent, but completely
anticorrelated sources. We have

\begin{equation}
(I_{L},\, I_{R})=(1,0)\,\textrm{or}\,(0,1),\end{equation}
each with probability $1/2$ (in every {}``elementary timestep'').
$\left\langle I_{3}\right\rangle $ and $\left\langle \delta I_{3}^{2}\right\rangle $
give the same result as before, Eq. (\ref{eq:simpleclass}).

We will now apply the dephasing terminal calculation to our simplified
model, see Fig. \ref{cap:dephclassical}d. 

Following closely the procedure of the previous section, we will at
first calculate the shot noise at the output port without taking into
account the fluctuations of the distribution function in the dephasing
terminal. This corresponds to the calculation of the {}``intrinsic''
current fluctuations $\delta I$. In the previous section, this was
done using the scattering theory of shot noise, and it will be performed
on the basis of classical probability theory in this section. It means
the two inputs to beamsplitter $B$ are treated as uncorrelated sources
of electrons (see Fig. \ref{cap:dephclassical}c). At first sight
we expect this to give different results than before, possibly with
an increased shot noise at the output ports, as the shot-noise supression
due to anticorrelations is lifted. According to this expectation,
accounting for the anticorrelations (by way of the fluctuating distribution
function) would then decrease the shot noise result and ultimately
give the {}``correct'' answer, obtained in the previous paragraph.
Nevertheless, this will turn out \emph{not} to be the case, the shot
noise \emph{without} anticorrelations will turn out to be the same
as before (and the subsequent introduction of anticorrelations will
spoil the agreement with the correct result).

The complete model is now described by:

\begin{equation}
(I_{L},I_{R})=(1,0)\,\textrm{or}\,(0,1),\end{equation}
each with probability $1/2$ in every timestep, and, \emph{independently}, 

\begin{equation}
I_{\varphi,\textrm{out}}=1\,\textrm{or}\,0,\end{equation}
with probability $1/2$, for the current entering the second beam
splitter from the left arm (i.e. after the dephasing terminal). 

The average current is $1/2$, as before. However, we have to be careful
when calculating the shot noise, as two electrons might impinge simultaneously
onto $B$ (ellipse in Fig. \ref{cap:dephclassical}c), in which case
a classical treatment would permit both to go into the same output
port (with probability $T_{B}R_{B}$ in the present model), while
in reality the Pauli principle prevents them from doing so. We find
the following table of probabilities, each line occuring with probability
$1/4$:

\begin{center}\begin{tabular}{|c|c|c|c|c|}
\hline 
$I_{\varphi,out}$&
$I_{R}$&
$P(I_{3}=0)$&
$P(I_{3}=1)$&
$P(I_{3}=2)$\tabularnewline
\hline
\hline 
0&
0&
1&
$0$&
$0$\tabularnewline
\hline 
1&
0&
$T_{B}$&
$R_{B}$&
$0$\tabularnewline
\hline 
0&
1&
$R_{B}$&
$T_{B}$&
$0$\tabularnewline
\hline 
1&
1&
0&
$1$&
0\tabularnewline
\hline
\end{tabular}\end{center}

From this, we obtain

\begin{equation}
\left\langle \delta I_{3}^{2}\right\rangle =\frac{1}{4}\,,\end{equation}
which happens to be identical to the result calculated for anticorrelated
inputs. If, however, we had neglected the Pauli principle, we would
have obtained a larger shot noise,

\begin{equation}
\left\langle \delta I_{3}^{2}\right\rangle =\frac{1}{4}+\frac{1}{2}R_{B}T_{B}\,(\textrm{no Pauli principle)}.\end{equation}
Therefore, the inclusion of the Pauli principle effects at the second
beamsplitter has suppressed the shot noise by $R_{B}T_{B}/2$. 

However, according to the logic of the dephasing terminal approach,
we still have to ensure the total current into the dephasing terminal
to vanish at each point in time. We will proceed as for the full dephasing
terminal calculation of the preceding section, i.e. by postulating
a fluctuating distribution function at the terminal which is chosen
to compensate the fluctuations $\delta I_{\varphi}$ that would be
present otherwise. Although this will effectively (and correctly)
re-introduce some anti-correlations between the two input currents
to the beamsplitter $B$, the postulated relation between {}``intrinsic''
current fluctuations $\delta I_{\varphi}$ and corresponding distribution
function fluctuations $\delta f_{\varphi}$ (Eq. (\ref{eq:dephdistr}))
constitutes an ad hoc semiclassical \emph{ansatz}. This is in contrast
to the rest of the dephasing terminal approach, which just represents
a valid model of a particular scattering geometry, designed to mimick
some aspects of dephasing. 

As a consequence, the full current fluctuations in the output port
are changed (see Section \ref{sec:Dephasing-terminal-approach}):

\begin{equation}
\Delta I_{3}=\delta I_{3}+R_{B}\delta I_{\varphi}\label{eq:extrafluct}\end{equation}
We will now calculate $\left\langle \Delta I_{3}^{2}\right\rangle $
by taking the correlators $\left\langle \delta I_{3}\delta I_{\varphi}\right\rangle $,
$\left\langle \delta I_{3}^{2}\right\rangle $ and $\left\langle \delta I_{\varphi}^{2}\right\rangle $
from the underlying classical model, instead of the quantum mechanical
scattering theory of shot noise. Since $\left\langle \delta I_{3}^{2}\right\rangle =1/4$
alone would give the correct result for the noise of the output current
(see above), it is already clear at this point that any further contributions
must lead to an artificial deviation from the correct value. 

The total current into the dephasing terminal is

\begin{equation}
I_{\varphi}=I_{L}-I_{\varphi,\textrm{out}},\end{equation}
which is forced to be zero at all times. Using this relation, as well
as the probabilities prescribed above, we find

\begin{equation}
\left\langle \delta I_{\varphi}^{2}\right\rangle =1/2\end{equation}
and

\begin{equation}
\left\langle \delta I_{3}\delta I_{\varphi}\right\rangle =-1/4.\end{equation}
This finally gives:

\begin{equation}
\left\langle \Delta I_{3}^{2}\right\rangle =\frac{1}{4}-\frac{1}{2}R_{B}T_{B}.\label{eq:lowsn}\end{equation}

Therefore, the shot noise calculated with the help of the dephasing
terminal ansatz is reduced (at $T_{B}\neq0,1$) as compared to what
is found for the original model. The reason should have become transparent
from our step-by-step derivation: The ansatz (\ref{eq:extrafluct})
serves to (correctly) take into account anticorrelations between the
two inputs to beamsplitter $B$, but it does \emph{not} throw out
the Pauli principle effects that determine the shot noise result for
two uncorrelated sources. In reality, only one effect or the other
is present, while the dephasing terminal approach keeps both of them,
thereby artificially reducing the shot noise. In the full calculation
of Section \ref{sec:Dephasing-terminal-approach}, the problem can
be traced to the ansatz describing the fluctuations $\delta f_{\varphi}$
of the distribution function as a fluctuating c-number function of
time. In that way, the dephasing terminal approach is no longer fully
quantum-mechanical (in contrast to the calculation of the current
itself, where $\delta f_{\varphi}$ is not needed).

It is interesting to note that there is no problem if we assume the
path-length difference between the two arms of the interferometer
to be large ($eVv_{F}\delta x\gg1$). We can incorporate this within
the simple classical model by assuming there to be a time-lag between
the anticorrelated input streams to the second beam splitter. Going
through steps similar to those above, we find a shot-noise reduction
even in the initial classical model, to a value given by Eq. (\ref{eq:lowsn}),
which is also the value found from the full dephasing terminal calculation
for that limit. This fits perfectly with our reasoning from above:
In this case, the anticorrelations and the effects due to the Pauli
principle are indeed present at the same time.

In this section we have demonstrated that the ansatz used for shot
noise calculations in presence of the dephasing terminal fails to
give the correct result when applied to this simple model of incoherent
transmission through a Mach-Zehnder interferometer. Moreover, the
(artificially reduced) shot noise result is even identical to what
is found in the more sophisticated calculation (Section \ref{sec:Dephasing-terminal-approach}),
where the dephasing terminal ansatz is the same but correlators are
evaluated using the scattering theory of shot noise (instead of simple
classical probabilities). Thus it is likely that this calculation
is affected by the same problem, which artificially reduces shot noise
(and makes it coincide with the phase-averaged shot-noise result). 

Strictly speaking, we have not given a direct proof of the failure
of the dephasing terminal ansatz for shot noise calculations, as we
have been able to present a detailed analysis only by taking the heuristic
classical model as our starting point. This has been necessary because
we lack a simple quantum-mechanical version of the fully incoherent
case, against which we should compare the results of the dephasing
terminal. Although we do consider a microscopic model in the next
section, the results obtained there cannot necessarily be compared
with the dephasing terminal either, since the dephasing terminal represents
a phenomenological model and it is unclear to which miscroscopic models
it should correspond (if any). Nevertheless, the arguments of this
section strongly suggest that the results of the dephasing terminal
approach to shot noise should be treated with caution, at least for
geometries similar to the two-way interferometer considered in this
article.

\section{Dephasing by classical noise}

\label{sec:Dephasing-by-classical}

\begin{figure}
\includegraphics[%
  width=3in,
  keepaspectratio]{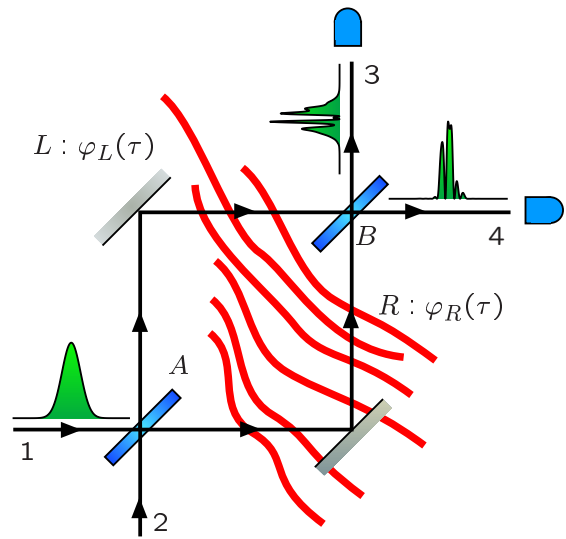}

\caption{\label{cap:fig1}The Mach-Zehnder interferometer setup analyzed in
the text. In the case shown here, the fluctuations of the environment
are fast compared with the temporal extent of the wave packet (determined
by temperature or voltage, see text). The probability density of the
incoming wave packet and its two outgoing parts is shown.}
\end{figure}
In this section, we will introduce a microscopic, fully quantum-mechanical
model of dephasing, and derive the resulting current-noise. Its major
advantages are that it displays the dependence of the results on the
power spectrum of the environmental fluctuations (which cannot be
done in any of the phenomenological models discussed above), that
it may be related directly to microscopic fluctuations acting on the
electrons, and that it properly treats all quantum-mechanical effects
regarding the motion of electrons. A brief discussion of the model
and some of the most important results has already been presented
in Ref.~\onlinecite{thePRL}.

There is one major simplifying feature of the model we are going to
use: We will assume dephasing to be induced by the fluctuations of
a \emph{classical} potential $V(x,t)$, acting on the electrons as
they traverse the interferometer (see Fig. \ref{cap:fig1}). Classical
noise may be used as an approximation to the effects of a truly quantum-mechanical
environment (e.g. phonons, Nyquist noise due to other electrons in
nearby gates). In fact, this approximation has been employed in the
past, e.g. in the theory of dephasing in weak localization\cite{classnoise}.
The idea is to use a classical fluctuating potential whose correlator
is set equal to the symmetrized part of the quantum-mechanical correlation
function. However, the zero-point fluctuations are (usually) to be
omitted, since Golden-Rule type calculations suggest that their effect
is canceled by the restriction of the scattering phase space via the
Pauli principle. Nevertheless, this approximation of neglecting the
environmental zero-point fluctuations can only be good as long as
$eV\ll T$. Otherwise, the scattering phase space will be determined
by the nonequilibrium Fermi functions in the arms of the interferometer,
and thus this simple prescription must fail.

On the other hand, our model may also describe nonequilibrium (classical)
microwave noise impinging onto the interferometer setup, or some thermal
noise source that behaves essentially classically (i.e. where $\omega<T$
for the relevant fluctuation frequencies). In this case, the treatment
becomes exact and holds for all values of voltage and temperature.

As the noise is classical, we still have a (time-dependent) single-particle
problem, i.e. we can solve for the motion of individual electrons.
The Fermi function will enter only in the end, when expectation values
(such as current correlators) are calculated. The Pauli principle
does not enter the calculation (except for, possibly, the potential
correlator, as explained above). In contrast, for the case of a fully
quantum-mechanical environment, we would end up with a complicated
many-body problem, since in any case the electrons would feel an effective
interaction induced by the coupling to the bath - even if we were
to neglect their intrinsic interaction in the beginning. This problem
is deferred to a future analysis.

\subsection{Electron field at output port}

As the electrons travel along the interferometer arms, they will accumulate
a random phase, due to the fluctuating potential. We neglect the additional
effects of the potential, namely acceleration and decceleration, by
assuming the electron's velocity to remain constant (linearized dispersion
relation). This should be a good approximation for sufficiently large
Fermi energy. The effects of a non-constant velocity have been analysed
in more detail in Ref.~\onlinecite{seelig}, where Nyquist-noise
induced dephasing of the current in a Mach-Zehnder setup has been
studied using the WKB approximation (see also Ref.~\onlinecite{seelig2}).
There, the main contribution to the end-result for the dephasing rate
did not depend on these extra effects. We will also assume backscattering
to be absent (i.e. the electrons are traveling along chiral edge channels,
or the potential is sufficiently smooth to prevent $2k_{F}$ momentum
transfers). Finally, as we are taking a model of non-interacting electrons
as our starting point, the electrons' spin does not play any important
role (except for trivial factors), and we assume the electrons to
be spin-polarized in the following.

The Heisenberg equation of motion for the electron field $\hat{\Psi}$
moving at constant velocity $v_{F}$, under the action of a fluctuating
potential $V(x,t)$, reads: 

\begin{equation}
i\partial_{t}\hat{\Psi}(x,t)=\left[\epsilon_{F}+v_{F}(-i\partial_{x}-k_{F})+V(x,t)\right]\hat{\Psi}(x,t)\,.\end{equation}
Here $x$ denotes the coordinate along the respective arm of the interferometer.
By solving this equation of motion, and taking into account the action
of the beamsplitters, we arrive at the following expression for the
electron field at some point in the outgoing lead $3$:

\begin{equation}
\hat{\Psi}(x,\tau)=\int\frac{dk}{\sqrt{2\pi}}e^{-i\epsilon_{k}\tau}\sum_{\alpha=1}^{3}t_{\alpha}(k,\tau)\hat{a}_{\alpha}(k)e^{s_{\alpha}ik_{F}x}\label{eq:psi}\end{equation}
 The field $\hat{\Psi}$ is a linear superposition of the electron
fields $\hat{a}_{\alpha}$ emitted at the reservoirs $\alpha=1,2,3$.
Note that for the special case of chiral edge channels, we may choose
to concentrate only on the outgoing current, such that $\alpha=3$
would be absent from Eq. (\ref{eq:psi}), and the corresponding trivial
contributions to subsequent equations would drop out as well. We have
$t_{3}=1$, $s_{1,2}=1,\, s_{3}=-1$, the reservoir operators obey
$\left\langle \hat{a}_{\alpha}^{\dagger}(k)\hat{a}_{\beta}(k')\right\rangle =\delta_{\alpha\beta}\delta(k-k')f_{\alpha}(k)$
with $f_{\alpha}$ the distribution function in reservoir $\alpha$,
and the integration is over $k>0$ only.

In contrast to the usual case, the transmission amplitudes $t_{\alpha}$
have become time-dependent. The amplitudes $t_{1},\, t_{2}$ for an
electron to go from terminal $1$ or $2$ to the output terminal $3$
depend on fluctuating time-dependent phases $\varphi_{L,R}$:

\begin{eqnarray}
t_{1}(k,\tau) & = & t_{A}t_{B}e^{i\varphi_{R}(\tau)}+r_{A}r_{B}e^{i\varphi_{L}(\tau)}e^{i(\phi+k\delta x)}\label{eq:t1}\\
t_{2}(k,\tau) & = & t_{A}r_{B}e^{i\varphi_{L}(\tau)}e^{i(\phi+k\delta x)}+r_{A}t_{B}e^{i\varphi_{R}(\tau)}\label{eq:t2}\end{eqnarray}
Here $t_{A/B}$ and $r_{A/B}$ are energy-independent transmission
and reflection amplitudes at the two beamsplitters (with $t_{j}^{*}r_{j}=-t_{j}r_{j}^{*}$),
$\delta x$ accounts for a possible path-length difference between
the interferometer arms, and $\phi$ denotes the Aharonov-Bohm phase
due to the flux through the interferometer. The electron accumulates
fluctuating phases while moving along the left or right arm: 

\begin{equation}
\varphi_{L,R}(\tau)=-\int_{-\tau_{L,R}}^{0}dt'\, V(x_{L,R}(t'),\tau+t')\,,\label{eq:phiV}\end{equation}
where $\tau$ is the time when the electron leaves the second beamsplitter
after traveling for a time $\tau_{L,R}$ along the interferometer
arms, the trajectories being described by $x_{L,R}(t)$. 

In our model, the total traversal times $\tau_{L,R}$ enter only at
this point, determining the relation between the phase correlator
and the potential correlator. Note that we have assumed the interaction
to be confined to the interferometer region. This assumption is natural
if the fluctuations are due to gates or other localized disturbances.
It is also sufficient for short-wavelength fluctuations. However,
in the case of long-wavelength fluctuations, it means that the effect
of these fluctuations on the phase difference $\varphi_{L}-\varphi_{R}$
will cancel out only in the case of vanishing path-length-difference.
Otherwise, cutting off the potential $V$ at the entry and exit beamsplitter
automatically introduces some remaining fluctuations in $\varphi_{L}-\varphi_{R}$.

In general, the form of the phase correlator can be related to the
potential correlator $\left\langle VV\right\rangle $, using Eq. (\ref{eq:phiV}).
For abbreviation, we set $V_{L}(t_{1},\tau)\equiv V(x_{L}(t_{1}),\tau+t_{1})\theta(-t_{1})\theta(t_{1}+\tau_{L})$
and likewise for $V_{R}$. Then we have $\varphi_{L}(\tau)=-\int dt'V_{L}(t',\tau+t')$
and thus:

\begin{eqnarray}
\left\langle \delta\varphi(\tau)\delta\varphi(0)\right\rangle =\int dt_{1}dt_{2}\left\langle \left(V_{L}(t_{1},\tau)-V_{R}(t_{1},\tau)\right)\right.\times\nonumber \\
\left.\left(V_{L}(t_{2},0)-V_{R}(t_{2},0)\right)\right\rangle \,.\label{eq:generalphiphi}\end{eqnarray}
The terms of the type $\left\langle V_{L}V_{L}\right\rangle $ and
$\left\langle V_{R}V_{R}\right\rangle $ describe phase fluctuations
within the two arms separately, while the cross-terms $\left\langle V_{L}V_{R}\right\rangle $
will serve to suppress dephasing in the case of long-wavelength fluctuations.
In a diagrammatic treatment of dephasing (e.g. in weak localization),
the cross-terms would correspond to {}``vertex contributions'',
whereas the former relate to {}``self-energy terms''.

In general, the potential correlator $\left\langle VV\right\rangle _{q\omega}$
and the corresponding phase correlator (Eq. \ref{eq:generalphiphi})
depend on the microscopic environment under consideration (cf. Ref.~\onlinecite{seelig}
for a calculation of spatially homogeneous potential fluctuations
in the interferometer arms, due to Nyquist noise), as well as the
geometry. A discussion of the potential and phase fluctuations for
realistic microscopic dephasing mechanisms will be provided in a future
work. Here we take the position that the phase correlator is given,
and we want to obtain the consequences for the current noise.

\subsection{Current}

Given the expression for the electron field, it is now in principle
straightforward to calculate the current and its correlators. In calculating
these quantities, we have to take both a quantum-mechanical expectation
value, as well as an average over the random process $V(x,t)$, or
rather $\varphi_{L,R}$. This average will be denoted by $\left\langle \cdot\right\rangle _{\varphi}$
in the following. The output current,

\begin{equation}
\hat{I}(\tau)=e\hat{\Psi}^{\dagger}(x,\tau)\frac{-i\partial_{x}}{2m}\hat{\Psi}(x,\tau)+h.c.,\end{equation}
follows from (\ref{eq:psi}). We will set $x=0$, as well as $\epsilon_{k'}-\epsilon_{k}=v_{F}(k'-k)$: 

\begin{eqnarray}
\hat{I}(\tau) & = & \frac{ev_{F}}{4\pi}\int dkdk'\, e^{i(\epsilon_{k'}-\epsilon_{k})\tau}\,\left[\sum_{\alpha}t_{\alpha}(k,\tau)\hat{a}_{\alpha}(k)\right]^{\dagger}\nonumber \\
 &  & \times\sum_{\beta}s_{\beta}t_{\beta}(k',\tau)\hat{a}_{\beta}(k')+h.c.\end{eqnarray}
Therefore, we have:

\begin{eqnarray}
I & = & \left\langle \left\langle \hat{I}\right\rangle \right\rangle _{\varphi}=\label{eq:curr}\\
 &  & \frac{ev_{F}}{2\pi}\int dk\,\left\{ -f_{3}(k)+\sum_{\alpha=1,2}f_{\alpha}(k)\left\langle T_{\alpha}(k)\right\rangle _{\varphi}\right\} \,.\nonumber \end{eqnarray}
The current depends on the phase-averages of transmission probabilities
$T_{1}=\left|t_{1}\right|^{2}$ and $T_{2}=1-T_{1}$: \begin{equation}
\left\langle T_{1}\right\rangle _{\varphi}=T_{A}T_{B}+R_{A}R_{B}+2z\left(r_{A}r_{B}\right)^{*}t_{A}t_{B}\cos(\phi+k\delta x),\label{eq:prob}\end{equation}
The interference term is suppressed by the factor $z\equiv\left\langle e^{i\delta\varphi}\right\rangle _{\varphi}$,
where $\delta\varphi=\varphi_{L}-\varphi_{R}$. In writing down this
expression, we have assumed $V(x,t)$ and thus $\delta\varphi$ to
be distributed symmetrically around $0$, such that $\left\langle \sin(\delta\varphi)\right\rangle _{\varphi}=0$.
For the special case of Gaussian statistics (which we will assume
below), we have $z=\exp(-\left\langle \delta\varphi^{2}\right\rangle /2)$.
The factor $z$ decreases the visibility of the interference pattern
observed in $I(\phi)$, and it has been defined to correspond precisely
to the phenomenological $z$ introduced for the dephasing terminal
model (see Eq. (\ref{eq:dephtermS})). An additional suppression of
the interference term may be brought about by the $k$-integration
in Eq. (\ref{eq:curr}), if $T\delta x/v_{F}>1$ or $eV\delta x/v_{F}>1$.
With respect to the current, it is indistinguishable from dephasing,
which provides the motivation of looking at shot noise in this context
(see Section \ref{sec:Shot-noise-as}). 

We note that the current itself is independent of the spectrum of
environmental fluctuations, as it only depends on the probability
distribution of $\delta\varphi$ at any given moment (and not its
time-dependent correlator). This will change when we look at shot
noise. It would also be different for the case of a quantum-mechanical
environment, where the {}``effective'' spread of $\delta\varphi$
would depend on the part of the bath spectrum that is still active
in dephasing, despite Pauli blocking.

\subsection{Noise power: General formula}

Before we turn to the calculation of the (zero-frequency) current
noise power $S$, we briefly list the main ingredients that we will
find below:

\begin{itemize}
\item A {}``classical current noise'' $S_{\textrm{cl}}$, which is due
to the time-dependent fluctuations of the interferometer's conductance.
The resulting current fluctuations are linear in the applied voltage,
such that the corresponding noise power is quadratic in $V$.
\item For any fixed external noise power, there is a finite current noise
contribution $S_{\textrm{V=0}}$ even at $V=0$ and $T=0$, due to
the nonequilibrium radiation impinging on the system.
\item The remainder of the full current noise contains the usual quantum-mechanical
partition noise $\mathcal{T}(1-\mathcal{T})$, which will be modified
due to the presence of the dephasing potential. The form of this modification
depends on whether the fluctuations of the environment are {}``fast''
or {}``slow'' as compared to the time-scales set by voltage and
temperature.
\end{itemize}
The full current noise power $S$ can be split into two parts, by
rewriting the irreducible current correlator:

\begin{eqnarray}
S=\int d\tau\,\left\langle \left\langle \hat{I}(\tau)\hat{I}(0)\right\rangle \right\rangle _{\varphi}-\left\langle \left\langle \hat{I}(0)\right\rangle \right\rangle _{\varphi}^{2}=\nonumber \\
\int d\tau\,\left\langle \left\langle \hat{I}(\tau)\right\rangle \,\left\langle \hat{I}(0)\right\rangle \right\rangle _{\varphi}-\left\langle \left\langle \hat{I}(0)\right\rangle \right\rangle _{\varphi}^{2}+\nonumber \\
\int d\tau\,\left\langle \left\langle \hat{I}(\tau)\hat{I}(0)\right\rangle -\left\langle \hat{I}(\tau)\right\rangle \left\langle \hat{I}(0)\right\rangle \right\rangle _{\varphi}\label{eq:generalsplit}\end{eqnarray}
The first integral on the r.h.s. describes shot noise due to the temporal
fluctuations of the conductance, i.e. fluctuations of a classical
current $I(\tau)=\left\langle \hat{I}(\tau)\right\rangle $ depending
on time-dependent transmission probabilities. We denote its noise
power as $S_{\textrm{cl}}$. It rises quadratically with the total
current, as is known from $1/f$-noise in mesoscopic conductors\cite{kogan}. 

We now focus on the second integral, which will contain the modified
partition noise (among other contributions, such as a finite {}``Nyquist
noise'' $S_{V=0}$). It is evaluated by inserting (\ref{eq:psi})
and applying Wick's theorem (similar formulas appear in Ref.~\onlinecite{lesovik99}):

\begin{widetext}

\begin{equation}
\left\langle \left\langle \hat{I}(\tau)\hat{I}(0)\right\rangle -\left\langle \hat{I}(\tau)\right\rangle \left\langle \hat{I}(0)\right\rangle \right\rangle _{\varphi}=\left(\frac{ev_{F}}{2\pi}\right)^{2}\int dkdk'\,\sum_{\alpha,\beta=1,2,3}f_{\alpha}(k)(1-f_{\beta}(k'))\, K_{\alpha\beta}(\tau)e^{i(\epsilon_{k'}-\epsilon_{k})\tau}.\label{eq:wideformula}\end{equation}
\end{widetext}Here $K_{\alpha\beta}$ is a correlator of four transmission
amplitudes. We have $K_{33}=1$, $K_{3\alpha}=K_{\alpha3}=0$, and

\begin{equation}
K_{\alpha\beta}(\tau)\equiv\left\langle t_{\alpha}^{*}(k,\tau)t_{\beta}(k',\tau)t_{\alpha}(k,0)t_{\beta}^{*}(k',0)\right\rangle _{\varphi}\,,\label{eq:AmplCorr}\end{equation}
for $\alpha,\beta=1,2$.

\subsection{Limiting cases}

\label{sub:Limiting-cases}In order to understand the resulting expressions,
we will now derive two limiting forms, for a {}``fast'' and a {}``slow''
environment. We will assume that the phase correlator $\left\langle \delta\varphi(\tau)\delta\varphi(0)\right\rangle $
decays on some time-scale $\tau_{c}$, the correlation time of the
environment. Note that even for a non-exponential decay we can still
define a typical scale $\tau_{c}$, e.g. by demanding $\left\langle \delta\varphi(\tau_{c})\delta\varphi(0)\right\rangle =\left\langle \delta\varphi^{2}\right\rangle /2$.
Now this time has to be compared against the other timescales, $(eV)^{-1}$
and $T^{-1}$. These scales enter the current noise formula (\ref{eq:wideformula})
in the form of the Fermi functions, and they determine the $\tau$-range
of the oscillating exponential factor, after integration over $k$
and $k'$. We will assume for the moment that the $k$-dependence
of $K_{\alpha\beta}$ itself is unimportant (i.e. $\delta x$ is sufficiently
small), see below for a discussion of other cases. 

For $eV\tau_{c}\ll1$ and $T\tau_{c}\ll1$ ({}``fast environment''),
the major contribution of the integration comes from $|\tau|\gg\tau_{c}$,
where $K_{\alpha\beta}$ factorizes into

\begin{equation}
K_{\alpha\beta}(\tau)\approx K_{\alpha\beta}(\infty)\equiv\left|\left\langle t_{\alpha}^{*}(k,0)t_{\beta}(k',0)\right\rangle _{\varphi}\right|^{2}\,.\label{eq:Klimit}\end{equation}
Adopting this limiting value for $K_{\alpha\beta}$ at all times $\tau$
yields the noise power

\begin{eqnarray}
\frac{S_{\textrm{fast}}}{e^{2}v_{F}/2\pi} & = & \int dk\,\sum_{\alpha,\beta=1,2}f_{\alpha}(1-f_{\beta})\left|\left\langle t_{\alpha}^{*}t_{\beta}\right\rangle _{\varphi}\right|^{2}+\nonumber \\
 &  & \,\,\,\,\,\, f_{3}(1-f_{3}),\label{eq:Sfast}\end{eqnarray}
where we have set $f_{\alpha,\beta}=f_{\alpha,\beta}(k)$ and $t_{\alpha,\beta}=t_{\alpha,\beta}(k,0)$.
Note that this form of the shot noise for a {}``fast'' environment
is not equivalent to an expression of the kind $\left\langle \mathcal{T}\right\rangle _{\varphi}(1-\left\langle \mathcal{T}\right\rangle _{\varphi})$,
which we have obtained from a simple classical model (see the discussion
in Section \ref{sec:Phenomenological-classical-model}). The difference
between those two formulas can be evaluated in general, and we find:

\begin{equation}
\left|\left\langle t_{1}^{*}t_{2}\right\rangle _{\varphi}\right|^{2}-\left\langle T_{1}\right\rangle _{\varphi}(1-\left\langle T_{1}\right\rangle _{\varphi})=(z^{2}-1)R_{B}T_{B}\,.\label{eq:comparefast}\end{equation}
This means the partition noise for the {}``fast'' case is usually
reduced below the value found from the simple expression. Nevertheless,
we will discuss a certain special case where the simple formula is
indeed recovered, see below.

We can always write the full noise power as

\begin{equation}
S=S_{\textrm{fast}}+S_{\textrm{fluct}}+S_{\textrm{cl}}\,,\label{eq:S}\end{equation}
where $S_{\textrm{fluct}}$ denotes the remainder besides $S_{\textrm{fast}}$
and $S_{\textrm{cl}}$, i.e. $S_{\textrm{fluct }}$ is given by Eq.
(\ref{eq:wideformula})) with $K_{\alpha\beta}(\tau)-K_{\alpha\beta}(\infty)$
inserted in place of $K_{\alpha\beta}(\tau)$. It yields a contribution
to the Nyquist noise $S_{V=0}$ (see below), but apart from that it
becomes important only at larger $V,\, T$, where it will serve to
produce the crossover to the case of the {}``slow'' environment,
which we discuss now. 

In the other limiting case the $\tau$-integration is dominated by
$|\tau|\ll\tau_{c}$ ({}``slow environment''), and we can use $K_{\alpha\beta}(\tau)\approx K_{\alpha\beta}(0)$,
which yields

\begin{eqnarray}
\frac{S_{\textrm{slow}}}{e^{2}v_{F}/2\pi} & = & \int dk\,\left\langle (f_{1}T_{1}+f_{2}T_{2})(1-(f_{1}T_{1}+f_{2}T_{2}))\right\rangle _{\varphi}+\,\nonumber \\
 &  & \,\,\,\,\, f_{3}(1-f_{3}),\label{eq:Sslow}\end{eqnarray}
i.e. the phase-average of the usual shot noise expression (at $T=0$
the expression in brackets reduces to $\left\langle T_{1}(1-T_{1})\right\rangle _{\varphi}$). 

\begin{figure}
\includegraphics[%
  width=3in,
  keepaspectratio]{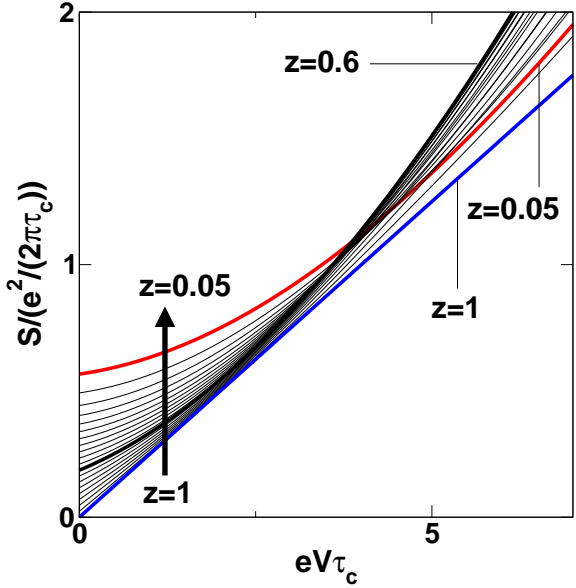}

\caption{\label{cap:fignoiseV}Full current noise $S$ as a function of $eV\tau_{c}$,
for increasing strength of dephasing ($z=1\ldots0.05$), according
to Eq. (\ref{eq:shotnoiseT0}). Dephasing always increases the current
noise beyond the value obtained for the ideal case $z=1$. The offset
$S_{V=0}$ is given in Eq. (\ref{eq:nyquist}), the slope near $V=0$
is described by $S_{\textrm{fast }}$, Eq. (\ref{eq:Sfast}), and
at higher voltages the dependence on $V$ is quadratic, due to $S_{\textrm{cl}}$.
When $S_{\textrm{cl}}$ is subtracted, the slope at large $eV\tau_{c}$
is determined by $S_{\textrm{slow}}$ (i.e. $\left\langle T_{1}(1-T_{1})\right\rangle _{\varphi}$).
Parameters: $T=0$, $\delta x=0$, $\phi=\pi/2$, $T_{A}=1/2$, $T_{B}=0.3$.}
\end{figure}

\subsection{Evaluation of shot noise contributions in general}

For a phase difference $\delta\varphi$ described by a Gaussian random
process of zero mean and prescribed correlator $\left\langle \delta\varphi(\tau)\delta\varphi(0)\right\rangle $,
the correlators $K_{\alpha\beta}$ of transmission amplitudes (Eq.
(\ref{eq:AmplCorr})) can be evaluated in general. This is done by
inserting the transmission amplitudes given above, and evaluating
the average of the exponential phase factors. We thus obtain an exact
expression which contains arbitrary orders of interaction with the
field (i.e. arbitrary powers of the phase correlator). 

\label{sec:Amplitude-correlators}The following expressions describe
the time-dependent deviation of the transmission amplitude correlators
$K_{\alpha\beta}(\tau)$ from their large-time limiting values $K_{\alpha\beta}^{\infty}$
(entering $S_{\textrm{fast}}$, Eq. (\ref{eq:Sfast})) . They follow
directly from the definition of $K_{\alpha\beta}(\tau)$, Eq. (\ref{eq:AmplCorr}),
using the transmission amplitudes in Eqs. (\ref{eq:t1}) and (\ref{eq:t2}),
as well as the abbreviations $g(\tau)=\exp\,\left\langle \delta\varphi(\tau)\delta\varphi(0)\right\rangle $
and $z=\exp(-\left\langle \delta\varphi^{2}\right\rangle /2)$:

\begin{eqnarray}
K_{12}(\tau)-K_{12}^{\infty}=K_{21}(\tau)-K_{21}^{\infty}=\nonumber \\
-2R_{A}T_{A}R_{B}T_{B}\cos(2\phi+\delta x(k+k'))z^{2}\left[g^{-1}(\tau)-1\right]\nonumber \\
+R_{B}T_{B}(R_{A}^{2}+T_{A}^{2})z^{2}\left[g(\tau)-1\right]\end{eqnarray}

\begin{eqnarray}
K_{11}(\tau)-K_{11}^{\infty}=K_{22}(\tau)-K_{22}^{\infty}=\nonumber \\
2R_{A}T_{A}R_{B}T_{B}z^{2}\nonumber \\
\times\left\{ \cos(2\phi+\delta x(k+k'))\left[g^{-1}(\tau)-1\right]+\left[g(\tau)-1\right]\right\} \end{eqnarray}

Here $R_{A}=1-T_{A}$, and we have repeatedly used the fact that there
is a phase shift of $\pm\pi/2$ between transmission and reflection
amplitudes at each beamsplitter ($r_{A}t_{A}^{*}=-r_{A}^{*}t_{A}$).

Both $S_{\textrm{fluct }}$ and $S_{\textrm{cl}}$ depend on the frequency
spectrum of the environment via the exponential $g(\tau)$ of the
phase correlator appearing in $K_{\alpha\beta}$ (in contrast, $S_{\textrm{fast}}$
and $S_{\textrm{slow}}$ are expressed in terms of $z=\exp(-\left\langle \delta\varphi^{2}\right\rangle /2)$
only). The resulting noise power can be written in terms of the following
Fourier transforms (with $n=\pm1$):

\begin{equation}
\hat{g}_{n}(\omega)\equiv\int d\tau\, e^{i\omega\tau}[e^{n\left\langle \delta\varphi(\tau)\delta\varphi(0)\right\rangle }-1].\label{eq:gdef}\end{equation}
Note that the first term in brackets approaches $1$ for $|\tau|\rightarrow\infty$,
as the phase correlations decay. These functions are similar to those
appearing in the so-called {}``P(E)-theory'' of tunneling in a dissipative
environment\cite{ingoldnazarov,schoendittrich} as well as in the
{}``independent boson model''. 

Using the explicit forms of the correlators $K_{\alpha\beta}$, we
find $S_{\textrm{fluct }}$ to be equal to:

\begin{eqnarray}
S_{\textrm{fluct}}=\left(\frac{ev_{F}}{2\pi}\right)^{2}\,\int dkdk'\,\left[f_{1}(1-f_{2}')+f_{2}(1-f_{1}')\right]\nonumber \\
\times R_{B}T_{B}\left\{ (R_{A}^{2}+T_{A}^{2})z^{2}\hat{g}_{+}(v_{F}(k'-k))-\right.\nonumber \\
\left.2\cos(2\phi+\delta x(k+k'))R_{A}T_{A}z^{2}\hat{g}_{-}(v_{F}(k'-k))\right\} +\nonumber \\
\left[f_{1}(1-f_{1}')+f_{2}(1-f_{2}')\right]\nonumber \\
\times2z^{2}R_{A}T_{A}R_{B}T_{B}\left\{ \hat{g}_{+}(v_{F}(k'-k))+\right.\nonumber \\
\left.\cos(2\phi+\delta x(k+k'))\hat{g}_{-}(v_{F}(k'-k))\right\} \label{eq:timepartsn2}\end{eqnarray}

In a similar fashion, we can evaluate $S_{\textrm{cl}}$. This term
does not display two different limiting regimes. The reason is that
it involves only correlators of time-dependent transmission probabilities,
but no oscillating factor depending on the energy difference. Therefore,
the result does not depend on the relation between $\tau_{c}$ and
$eV,\, T$. In general, this term is determined by the zero-frequency
correlators of the exponential phase factors contained in the transmission
probabilities. We find (with $\delta f\equiv f_{1}-f_{2}$):

\begin{eqnarray}
S_{\textrm{cl}}=2z^{2}R_{A}R_{B}T_{A}T_{B}\left(\frac{e}{2\pi}\right)^{2}v_{F}^{2}\,\int dkdk'\,\delta f\delta f'\,\,\nonumber \\
\times\left[\hat{g}_{-}(0)\cos(2\phi+\delta x(k+k'))+\hat{g}_{+}(0)\cos(\delta x(k-k'))\right]\, & .\label{eq:Scl}\end{eqnarray}

\subsection{Current noise at $T=0$}

\label{sub:Current-noise-at}It still remains to evaluate the $k$-integrals
contained in the expressions (\ref{eq:timepartsn2}),(\ref{eq:Scl})
for $S_{\textrm{fluct}}$ and $S_{\textrm{cl}}$. In this section,
we will present and discuss explicit expressions for the case $T=0$,
$\delta xeV/v_{F}\ll1$, i.e. the case of pure dephasing without any
thermal smearing. According to the discussion at the beginning of
the present section, analyzing the zero-temperature limit invariably
means we adopt the picture of real classical noise impinging onto
the system (as opposed to classical noise being an approximation for
a quantum bath, which would require $eV\ll T$ for selfconsistency).

We assume the electrons to be injected from reservoir $1$, i.e. $f_{2}(k)=f_{3}(k)=f(k)=\theta(k_{F}-k)$
and $f_{1}(k)=\theta(k_{F}+\Delta k-k)\equiv f(k)+\delta f(k)$, with
$\Delta k=eV/v_{F}$.

As we are interested in the \emph{shot noise}, we subtract the equilibrium
part $S_{\textrm{fluct}}(V=0)\equiv S_{V=0}$ from $S_{\textrm{fluct}}$
(Eq. (\ref{eq:timepartsn2})). In the remainder, the term stemming
from $f_{1}(1-f_{1}')$ is seen not to contribute (employing symmetry
in $k$ and $k'$), and the terms from $f_{1}(1-f_{2}')$ and $f_{2}(1-f_{1}')$
lead to the integral

\begin{eqnarray}
\int dkdk'\,(\delta f(1-f')-f\delta f')\,\hat{g}_{n}(v_{F}(k'-k))\nonumber \\
=\frac{2eV}{v_{F}^{2}}I_{n}(V),\label{eq:powerconvol}\end{eqnarray}
where $I_{n}(V)$ also depends on temperature $T$. In particular,
at $T=0$, we find $I_{n}$ to be: 

\begin{eqnarray}
I_{n}(V) & \equiv & \int_{0}^{eV}d\omega\,(1-\frac{\omega}{eV})\hat{g}_{\lambda}(\omega)\label{eq:Idef}\end{eqnarray}
Collecting the contributions from $S=S_{\textrm{cl}}+S_{\textrm{fast}}+S_{\textrm{fluct}}$,
the shot noise is then given by:

\begin{eqnarray}
\frac{S-S_{V=0}}{e^{3}V/2\pi} & = & \frac{eV}{\pi}\, z^{2}R_{A}R_{B}T_{A}T_{B}(\cos(2\tilde{\phi})\hat{g}_{-}(0)+\hat{g}_{+}(0))\nonumber \\
 &  & +\left|\left\langle t_{1}^{*}t_{2}\right\rangle _{\varphi}\right|^{2}\nonumber \\
 &  & +\frac{1}{\pi}\, z^{2}R_{B}T_{B}\left\{ -2\cos(2\tilde{\phi})R_{A}T_{A}\, I_{-}(V)\right.\nonumber \\
 &  & \left.+(R_{A}^{2}+T_{A}^{2})\, I_{+}(V)\right\} \label{eq:shotnoiseT0}\end{eqnarray}
Here we have defined the average phase as $\tilde{\phi}=\phi+k_{F}\delta x$.
The first line of Eq. (\ref{eq:shotnoiseT0}) corresponds to $S_{\textrm{cl}}$,
the second to $S_{\textrm{fast}}$, and the rest to $S_{\textrm{fluct}}-S_{V=0}$.
The current noise displayed in Eq. (\ref{eq:shotnoiseT0}) is a function
of $eV\tau_{c}$, $z$, $T_{A}$, $T_{B}$, $\phi$, and of the detailed
shape of the environment correlator contained in $I_{n}(V)$ and $\hat{g}_{n}(0)$.
The dependence of $S-S_{V=0}$ on voltage is explicit in the first
two lines, stemming from $S_{\textrm{cl}}$ and $S_{\textrm{fast}}$
(quadratic and linear, respectively). Only the contribution from $S_{\textrm{fluct}}$
(last two lines) depends on voltage in a more complicated way, via
the environment spectrum. 

We can introduce the dependence on the environment correlation time
$\tau_{c}$ by assuming the phase-correlator to be given as $\left\langle \delta\varphi(\tau)\delta\varphi(0)\right\rangle =C(\tau/\tau_{c})$.
Then $I_{n}(V)$ is a function of $eV\tau_{c}$ only. 

We may confirm directly that $S_{\textrm{fast}}$ dominates at low
voltages, since $S_{\textrm{cl}}$ is quadratic in voltage and the
integrals $I_{\pm}(V)$ in $S_{\textrm{fluct}}$ vanish. At large
$eV\tau_{c}\gg1$ we can use the sum-rule

\begin{equation}
I_{n}(V)\rightarrow\pi\left[z^{-2n}-1\right]\label{sumrule}\end{equation}
 to combine the shot noise contributions in the last three lines of
Eq. (\ref{eq:shotnoiseT0}), i.e. $S_{\textrm{fast}}+S_{\textrm{fluct}}-S_{\textrm{V=0}}$,
yielding:

\begin{eqnarray}
\left|\left\langle t_{1}^{*}t_{2}\right\rangle _{\varphi}\right|^{2}-2\cos(2\tilde{\phi})R_{A}T_{A}R_{B}T_{B}\, z^{2}(z^{2}-1)\nonumber \\
+R_{B}T_{B}(R_{A}^{2}+T_{A}^{2})\,(1-z^{2})=\left\langle T_{1}(1-T_{1})\right\rangle _{\varphi}\label{eq:combined}\end{eqnarray}
This is precisely the result expected from the limit of a {}``slow''
bath, i.e. from $S_{\textrm{slow}}$, compare Eq. (\ref{eq:Sslow}).
At intermediate voltages, the shot noise interpolates smoothly between
the extremes described by $S_{\textrm{fast}}$ and $S_{\textrm{slow}}$.
\begin{figure}
\includegraphics[%
  width=0.95\columnwidth]{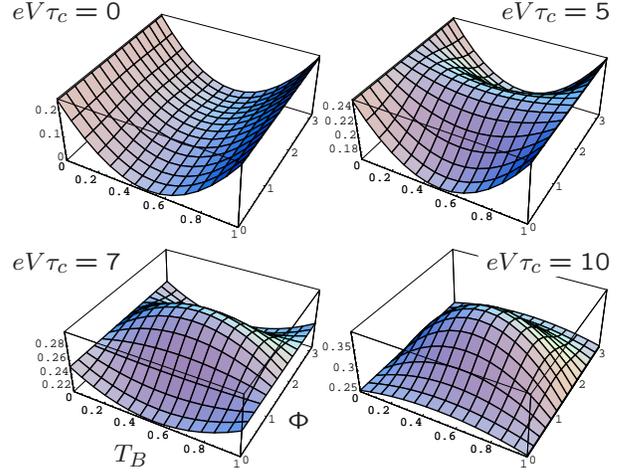}

\caption{\label{figSNze}Normalized shot noise $(S-S_{V=0})/(e^{3}V/2\pi)$
of the Mach-Zehnder interferometer, as a function of the transmission
of the second beamsplitter, $T_{B}$ (horizontal axis), and the phase
difference $\phi$ (into the plane), for the case of small but finite
visibility, $z=1/e$, at $T=\delta x=0$ and $T_{A}=1/2$. The different
plots show the succession from a {}``fast'' environment to a {}``slow''
one, by increasing the voltage or the correlation time (top left to
bottom right: $eV\tau_{c}=0,\,5,\,7,\,10$). Note the change of plot
range on the vertical axis. At $T_{B}=0,1$, the normalized shot noise
remains fixed at $1/4$.}
\end{figure}

To produce the plots discussed in the following, we have assumed 50\%
transparency of the first beamsplitter ($T_{A}=1/2$) and a simple
Gaussian form for the phase correlator,

\begin{equation}
\left\langle \delta\varphi(\tau)\delta\varphi(0)\right\rangle =\left\langle \delta\varphi^{2}\right\rangle e^{-(\tau/\tau_{c})^{2}}\,.\end{equation}

In the case of $T_{A}=1/2$, the normalized shot noise is given explicitly
as a function of the parameters $z,eV\tau_{c},T_{B},\tilde{\phi}$
by the following formula:

\begin{eqnarray}
\frac{S-S_{V=0}}{e^{3}V/2\pi} & = & \frac{z^{2}}{4\pi}(eV\tau_{c})R_{B}T_{B}(\cos(2\tilde{\phi})\,\hat{\tilde{g}}_{-}(0)+\hat{\tilde{g}}_{+}(0))+\nonumber \\
 &  & +\frac{1}{4}\left[(T_{B}-R_{B})^{2}+4z^{2}R_{B}T_{B}\sin^{2}\tilde{\phi}\right]\nonumber \\
 &  & +\frac{z^{2}}{2\pi}R_{B}T_{B}[\tilde{I}_{+}(V\tau_{c})-\cos(2\tilde{\phi})\tilde{I}_{-}(V\tau_{c})]\,.\label{eq:snexpl}\end{eqnarray}
Here the functions $\hat{\tilde{g}}_{n}$ and $\tilde{I}_{n}$ are
evaluated by setting $\tau_{c}=1$ in the phase correlator $C(\tau/\tau_{c})$.
They have to be evaluated by numerical integration (for a given shape
$C(\tau/\tau_{c})$ of the phase correlator). This equation has been
derived for the case $eV>0$, but it may be verified that $S$ is
symmetric in $V$.

The full current noise $S$ also contains the Nyquist noise $S_{V=0}$,
which is independent of $\tilde{\phi}$ and $T_{A}$: 

\begin{equation}
S_{V=0}=\frac{e^{2}}{2\pi^{2}}z^{2}R_{B}T_{B}\int_{0}^{\infty}d\omega\,\omega\hat{g}_{+}(\omega)\,.\label{eq:nyquist}\end{equation}
The Nyquist noise scales like $1/\tau_{c}$. The dependence on $z$
is not explicit, as the integral depends on $z$ itself (scaling like
$1/z^{2}$ for small but not ultrasmall $z$). In deriving the Nyquist
noise from $S_{\textrm{fluct}}$, we have only kept the contribution
from states near the Fermi edge, assuming all states for $k\in(-\infty,k_{F})$
to be filled. 

Fig. \ref{cap:fignoiseV} shows the evolution of $S(V)$ with increasing
dephasing strength (i.e. increasing $\left\langle \delta\varphi^{2}\right\rangle $,
decreasing $z$). Note that the shot noise itself (i.e. the deviation
from $V=0$) may even vanish due to the presence of the fluctuating
potential, in the limit of a {}``fast'' environment, $V\tau_{c}\rightarrow0$:
According to Eq. (\ref{eq:Sfast}), $S_{\textrm{fast}}$ is determined
by $\left|\left\langle t_{1}^{*}t_{2}\right\rangle _{\varphi}\right|^{2}$
at $T=0$. In the limit of vanishing visibility, $z\rightarrow0$,
this expression is zero for $T_{B}=1/2$, independent of the value
of $T_{A}$. That may be verified explicitly, but it can also be deduced
from Eq. (\ref{eq:comparefast}), by noting that $\left\langle T_{1}\right\rangle _{\varphi}(1-\left\langle T_{1}\right\rangle _{\varphi})=1/4$
for $z=0,\, T_{B}=1/2$. However, although $S_{\textrm{fast}}$ can
become zero, the total current noise $S$ does not vanish, due to
the Nyquist contribution (and the classical term $S_{\textrm{cl }}$
at higher voltages). Indeed the figure illustrates that the fluctuating
potential $V(x,t)$ always leads to an increase in current noise (as
expected). Nevertheless, the dependence on dephasing strength may
be non-monotonic, as seen in Fig. \ref{cap:fignoiseV}, at large voltages
$V$. 

The dependence on $eV\tau_{c}$ is also illustrated in Fig. \ref{figSNze},
where the dependence of the shot noise on the parameters $T_{B}$
and $\phi$ is displayed for different values of $eV\tau_{c}$ (see
also the figures in Ref.~\onlinecite{thePRL} showing the crossover
between $S_{\textrm{fast}}$ and $S_{\textrm{slow}}$). 

Note that the behaviour of $S_{\textrm{fast}}$, given by $\left|\left\langle t_{1}^{*}t_{2}\right\rangle _{\varphi}\right|^{2}$,
is quite different from that of $\left\langle T_{1}\right\rangle _{\varphi}(1-\left\langle T_{1}\right\rangle _{\varphi})$,
which is the form derived from the simple classical model of Section
\ref{sec:Phenomenological-classical-model}. Indeed, the latter expression
does not vanish at intermediate values of $T_{A},\, T_{B}$ ($\neq0,1$),
and for $z=0$ it becomes independent of $T_{B}$ if $T_{A}=1/2$
(while the first expression becomes independent of $T_{A}$ if $T_{B}=1/2$).

\subsection{Other cases: Finite temperatures and finite path-length difference}

\label{sub:Other-cases:-Finite}The results of the previous section
have been derived for the case $T=0,\,\delta x=0$. We will now discuss
the changes introduced by relaxing these assumptions.

\emph{Finite temperatures:} If we calculate the current noise for
a finite temperature $T$, but still at $\delta x=0$, the different
components of $S=S_{\textrm{cl}}+S_{\textrm{fast}}+S_{\textrm{fluct}}$
show the following behaviour: The contrast of the current $I(\phi)$
is unaffected by the thermal smearing of the Fermi surfaces (since
$\delta x=0$), and for the same reason the {}``classical'' part
$S_{\textrm{cl}}$ remains the same (apart from possible changes related
to a temperature-dependence of the environmental power spectrum).
In $S_{\textrm{fast}}$ from Eq. (\ref{eq:Sfast}), the finite-temperature
Fermi functions lead to Nyquist noise contributions (which have been
absent in $S_{\textrm{fast}}$ for $T=0$):

\begin{eqnarray}
\frac{S_{\textrm{fast}}}{e^{2}/2\pi} & = & T\,\left\{ \left|\left\langle T_{1}\right\rangle _{\varphi}\right|^{2}+\left|\left\langle T_{2}\right\rangle _{\varphi}\right|^{2}+1\right\} +\nonumber \\
 &  & +eV\,\left|\left\langle t_{1}^{*}t_{2}\right\rangle _{\varphi}\right|^{2}\coth\left(\frac{\beta eV}{2}\right)\end{eqnarray}

In $S_{\textrm{fluct}}$ of Eq. (\ref{eq:timepartsn2}) the $k,k'$-integral
over products of Fermi functions and environment power spectra $\hat{g}_{\lambda}$
are altered as well. In the particular limit of $T\tau_{c}\rightarrow\infty$
(regardless of $V$), the Fermi functions can be approximated by a
constant on the scale over which the power spectrum $\hat{g}_{\lambda}$
changes. Then the integrals over $k'-k$ can be carried out easily,
leading to sum rules. Combining the terms from $S_{\textrm{fast}}$
and $S_{\textrm{fluct}}$ in this limit leads to the expression $S_{\textrm{slow}}$
(Eq. (\ref{eq:Sslow})). We conclude that $S_{\textrm{slow}}$ is
indeed the appropriate expression for $1/\tau_{c}\ll\textrm{max}(T,eV)$. 

\emph{Finite path-length difference}: If a finite path-length difference
$\delta x$ is introduced, we have to consider four time-scales altogether:
$\tau_{c}$, $(eV)^{-1}$, $T^{-1}$ and the new time-scale $\delta x/v_{F}$.
We will not give an exhaustive discussion of all possible cases for
the order of these times. In the limiting case of very small $\delta x$,
i.e. $\delta x/v_{F}\ll\tau_{c},(eV)^{-1},T^{-1}$, the previous expressions
remain unchanged. Even if $\delta x/v_{F}$ becomes larger than $\tau_{c}$
(but remains much smaller than $(eV)^{-1},T^{-1}$), it may still
be shown that this does not affect the results for the current noise.

We now consider the more interesting opposite limit, where the averaging
over wave-number $k$ is so important that it destroys completely
the interference pattern, i.e. $\delta x/v_{F}\gg(eV)^{-1}$ or $\delta x/v_{F}\gg T^{-1}$.
In that case, the interference term in the average current is completely
suppressed, such that the additional dephasing effect of the environment
is unimportant for the current. In addition, the {}``classical''
current noise part $S_{\textrm{cl}}$ now vanishes, since it depends
on the temporal fluctuation of the interference term in the average
current $\left\langle \hat{I}(\tau)\right\rangle $, which is already
absent due to thermal averaging. The other two parts $S_{\textrm{fast}}$
and $S_{\textrm{fluct}}$ of the current noise $S$ are changed as
well, but they do not become equal to the results obtained without
dephasing. 

We illustrate those changes in the zero-temperature case analyzed
in section \ref{sub:Current-noise-at}. The shot noise in Eq. (\ref{eq:snexpl})
is changed in the following ways: The first line (due to $S_{\textrm{cl}}$)
is absent, and the second and third lines (due to $S_{\textrm{fast}}$
and $S_{\textrm{fluct}}$) are averaged over the phase $\phi$, such
that the average of $\cos(2\phi)$ vanishes and that of $\sin^{2}\phi$
is equal to $1/2$. However, the shot noise still depends on $z$
and on the bath spectrum (via $I_{+}$):

\begin{eqnarray}
\frac{S-S_{V=0}}{e^{3}V/2\pi} & = & T_{A}R_{A}(T_{B}-R_{B})^{2}+\nonumber \\
 &  & z^{2}T_{B}R_{B}(T_{A}^{2}+R_{A}^{2})[1+\frac{I_{+}(V)}{\pi}]\,.\label{eq:largedx}\end{eqnarray}
For a {}``fast'' environment, we have $I_{+}(V)\rightarrow0$, such
that Eq. (\ref{eq:largedx}) becomes $T_{A}R_{A}(T_{B}-R_{B})^{2}$
in the fully incoherent case, $z\rightarrow0$. In the opposite limit
of large voltages ({}``slow environment'', $eV\tau_{c}\gg1$), we
have $I_{+}(V)\rightarrow\pi\left[z^{-2}-1\right]$, which makes Eq.
(\ref{eq:largedx}) independent of $z$. The resulting expression
is then equivalent to the one obtained by pure $k$-averaging, in
the absence of dephasing. In conclusion, shot noise may indeed help
to reveal the presence or absence of dephasing even when thermal averaging
is so strong that interference is already completely suppressed, but
not in the limit of a {}``slow'' bath. The Nyquist noise is not
affected by $\delta x$, since it results from setting $f_{1}=f_{2}$
in Eq. (\ref{eq:timepartsn2}), whence the $\cos$-terms depending
on $\delta x$ combine to zero. 

\emph{Beam of electrons}: It is instructive to notice that even the
dephasing model considered here can lead to the simple form $\left\langle T_{1}\right\rangle _{\varphi}(1-\left\langle T_{1}\right\rangle _{\varphi})$
of the shot-noise (which holds for a classical model in the fully
incoherent limit, see Section \ref{sec:Phenomenological-classical-model}).
This is true provided the transport situation is different from the
usual one treated above. Instead of having all the reservoirs filled
up to some Fermi level and then applying a voltage between them, we
consider a situation where a (nearly mono-energetic) beam of electrons
is injected from reservoir $1$, with wavenumbers in an interval $\Delta k$,
and all the other reservoirs are \emph{empty}: $f_{1}(k)=\theta(k\in[k_{F},k_{F}+\Delta k]$
and $f_{2}=f_{3}=0$. In this situation, there is no {}``Nyquist
noise'' ($S$ vanishes for $\Delta k=0$, when there are no electrons
at all). In the limit of small $\Delta k$ ({}``fast environment'',
with $\Delta kv_{F}\tau_{c}\ll1$), we obtain for the shot noise (at
$T=0$):

\begin{equation}
S-S_{\textrm{cl}}\approx\frac{e^{2}v_{F}}{2\pi}\Delta k\left\langle T_{1}\right\rangle _{\varphi}(1-\left\langle T_{1}\right\rangle _{\varphi})\,.\label{eq:snbeam}\end{equation}
Here we assumed $\delta x=0$ as above. This formula follows by evaluating
$S_{\textrm{fast}}+S_{\textrm{fluct}}$ in the limit of small $\Delta k$
and using the sum rule (\ref{sumrule}). In contrast to the evaluation
of $S_{\textrm{fluct}}$ in the transport situation considered above,
the integral over $k'$ now runs over all states and is not restricted
to a small transport window, which is essential to obtain (\ref{eq:snbeam}).
We conclude that this is yet another example\cite{gavish} of a situation
where the correct result for the shot noise cannot be obtained by
taking into account only the {}``surplus'' electrons in the transport
window of size $eV$, even though this approach \emph{does} yield
the correct current. The presence of the filled Fermi seas is not
merely important for the Nyquist noise contribution but for the shot
noise as well. In the other limiting case, $\Delta kv_{F}\tau_{c}\gg1$,
we obtain the result expected from $S_{\textrm{slow}}$, i.e. with
$\left\langle T_{1}(1-T_{1})\right\rangle _{\varphi}$ in Eq. (\ref{eq:snbeam}),
in addition to the {}``classical'' contribution $S_{\textrm{cl}}$
with its quadratic dependence on $\Delta k$.

\section{Comparison of different models and regimes}

\label{sec:Comparison-of-different}

In this section, we will collect and compare the various results obtained
for the different models (and different regimes). We will restrict
ourselves to the fully incoherent limit ($z=0$), at $T=0$ and $T_{A}=1/2$.
We emphasize, of course, that by comparing these models we do not
want to imply that one should expect them to agree in any limit. The
phenomenological classical model is simply a heuristic construction
that is known to give the correct result for a single barrier, and
even for the dephasing terminal ansatz it is not completely clear
to which microscopic model it is to correspond. In addition, we remind
the reader that the results obtained for dephasing by classical noise
are not expected to coincide with those obtained for a more elaborate
analysis of a quantum-mechanical bath, in the limit $eV\gg T$ considered
here.

\begin{figure}
\begin{center}\includegraphics[%
  width=0.90\columnwidth,
  keepaspectratio]{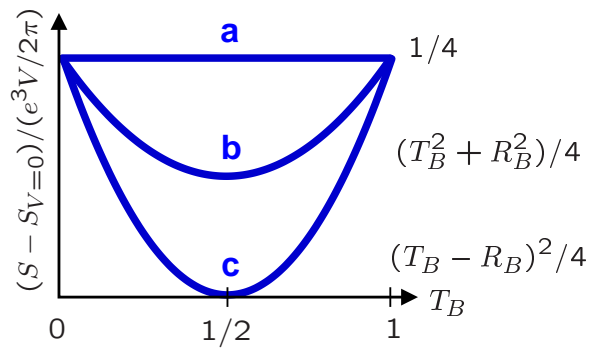}\end{center}

\caption{\label{cap:comparisonmodels}Shot noise as a function of the transmission
of the second beamsplitter, for the fully incoherent case (and $T_{A}=1/2$):
Different models and parameter regimes lead to different curves (see
text). }
\end{figure}
We have to distinguish three possible results for the modified partition
noise term, entering the \emph{shot-noise} $S-S_{V=0}$ (depicted
in Fig. \ref{cap:comparisonmodels}):

\begin{eqnarray}
 &  & (a)\,\left\langle T_{1}\right\rangle _{\varphi}(1-\left\langle T_{1}\right\rangle _{\varphi})=1/4,\nonumber \\
 &  & (b)\,\left\langle T_{1}(1-T_{1})\right\rangle _{\varphi}=(T_{B}^{2}+R_{B}^{2})/4,\nonumber \\
 &  & (c)\,\left|\left\langle t_{1}^{*}t_{2}\right\rangle _{\varphi}\right|^{2}=(T_{B}-R_{B})^{2}/4\label{eq:possibilities}\end{eqnarray}

The corresponding values for the different models and regimes are
indicated in the following table. Note that these expressions only
refer to the contribution to $S$ which is linear in voltage. For
the model of dephasing by classical noise (last three entries), one
still has to add the constant background $S_{V=0}$, as well as $S_{\textrm{cl}}$
(growing quadratically with voltage). 

\begin{center}\begin{tabular}{|l||c|c|}
\hline 
Model/regime&
$\delta x\ll v_{F}/eV$&
$\delta x\gg v_{F}/eV$\tabularnewline
\hline
\hline 
no dephasing ($z=1$)&
$T_{1}(\phi)(1-T_{1}(\phi))$&
b\tabularnewline
\hline 
Simple classical model&
a&
b\tabularnewline
\hline 
Dephasing terminal&
b&
b\tabularnewline
\hline 
{}``fast'' environment&
c&
c\tabularnewline
\hline 
{}``slow'' environment&
b&
b\tabularnewline
\hline 
{}``narrow electron beam''&
a&
a\tabularnewline
\hline
\end{tabular}\end{center}

For the particular parameters considered here, the presence of thermal
averaging ($\delta x\gg v_{F}/eV$) only affects the results obtained
without dephasing or from the simple classical model of Section \ref{sec:Phenomenological-classical-model}.
In any case, the results for the {}``slow classical noise'' (Eq.
(\ref{eq:Sslow})) and the dephasing terminal (Eq. (\ref{eq:shotnoise}))
both coincide with the result (b) $\left\langle T_{1}(1-T_{1})\right\rangle _{\varphi}$
obtained for complete thermal averaging (which is also obtained from
the simple classical model if thermal averaging is present on top
of dephasing). It might still be possible to deduce the presence of
dephasing in the case of classical noise, both from the presence $S_{V=0}$
and $S_{\textrm{cl}}$, although we have to note that $S_{\textrm{cl}}$
vanishes if both dephasing and thermal averaging are effective. The
form of the shot noise $S_{\textrm{fast }}$ (c) obtained in the limit
of a {}``fast'' environment (Eq. (\ref{eq:Sfast})) is not found
in any of the other models. Finally, the result (a) $\left\langle T_{1}\right\rangle _{\varphi}(1-\left\langle T_{1}\right\rangle _{\varphi})$
conjectured from the simple classical model (in the absence of thermal
averaging) can also be found for dephasing by classical noise, provided
we consider a special transport situation, with a {}``narrow beam
of electrons'' (Eq. (\ref{eq:snbeam})).

\section{Conclusions}

We have analyzed the effect of a fluctuating environment on the shot
noise in an electronic Mach-Zehnder interferometer. The environment
has been modeled as a classical noise field which leads to a fluctuating
phase difference for electrons traversing the interferometer and thereby
suppresses the interference term. For comparison, we have also discussed
a simple classial ansatz and the phenomenological dephasing terminal
approach.

The effect of dephasing on the average \emph{current} is always the
same, and qualitatively indistinguishable from {}``thermal averaging''
(averaging over wave number in the presence of a path-length difference).
However, important differences appear in the shot noise results. While
the power spectrum of the phase fluctuations does not enter the current
for the case of a classical fluctuating potential considered here,
the current \emph{noise} strongly depends on the fluctuation spectrum,
thus offering more information on the environment. There are three
main contributions to the current noise: some {}``classical'' current
noise (rising like $V^{2}$), due to the fluctuations of the conductance,
some {}``Nyquist noise'' background, and finally the usual partition
noise, modified due to the presence of the environment. The partition
noise contribution depends on a two-time correlator of four transmission
amplitudes and is sensitive to the power spectrum. We have distinguished
the limits of a {}``slow'' and a {}``fast'' environment, depending
on whether the inverse correlation time of fluctuations $1/\tau_{c}$
is much smaller or much larger than the maximum of voltage $eV$ and
temperature $T$. We have found that the usual result $T_{1}(1-T_{1})$
for the partition noise (at given transmission probability $T_{1}$)
may be replaced by one of three limiting forms, depending on the correlation
time $\tau_{c}$, the transport situation and the dephasing model:
(i) For a {}``slow'' environment, the usual result is averaged over
the phase fluctuations, $\left\langle T_{1}(1-T_{1})\right\rangle _{\varphi}$,
which is similar to the effect of thermal averaging and identical
to the result provided by the dephasing terminal (although there may
be problems with the dephasing terminal, see Section \ref{sec:Dephasing-terminal-approach}).
(ii) For a {}``fast'' environment applied to a nearly mono-energetic
beam of electrons, we obtain $\left\langle T_{1}\right\rangle _{\varphi}(1-\left\langle T_{1}\right\rangle _{\varphi})$,
which is also the result derived from a simple classical model. (iii)
For a {}``fast'' environment applied to the usual transport situation
(with the chemical potential of one of the input reservoirs increased
by $eV$), we obtain $\left|\left\langle t_{1}^{*}t_{2}\right\rangle _{\varphi}\right|^{2}$,
where $t_{1,2}$ are the amplitudes of reaching the output port from
inputs $1$ and $2$ ($\left|t_{1}\right|^{2}=1-\left|t_{2}\right|^{2}=T_{1}$).
In this case, the shot noise at $T=0$ can even be suppressed to zero
by the fluctuating environment for appropriate parameter combinations,
while on the other hand the Nyquist noise becomes nonzero.

We have discussed the crossover between {}``slow'' and {}``fast''
environment, the dependence of the shot noise on the phase difference
between the paths and on the beam splitter transparency, and the influence
of finite temperatures and finite path-length difference (thermal
averaging).

The most important tasks that remains to be tackled in future works
are the consideration of finite frequency shot noise, the derivation
of realistic microscopic power spectra as input for this calculation,
and, in particular, the inclusion of a truly quantum-mechanical environment,
which will be relevant particularly for the case of voltages larger
than temperature.

\section{Acknowledgments}

We thank M. Heiblum for useful comments and for sending us a preprint
of Ref.~\onlinecite{heiblum}. Useful discussions with W. Belzig,
M. Büttiker, A. Clerk, C. Egues, and E. Sukhorukov are gratefully
acknowledged. This work has been supported by the Swiss NSF and the
NCCR nanoscience, as well as DFG grant MA 2611/1-1.

\end{document}